\documentclass[aps,prd,showpacs,eqsecnum,twocolumn,superscriptaddress,nofootinbib]
{revtex4-1}
\usepackage{amsmath,amssymb,graphicx,color}

\begin{document}

\title{Sub-radian-accuracy gravitational waveforms 
  of coalescing binary neutron stars in numerical relativity}

\author{Kenta Kiuchi}\affiliation{Center for Gravitational Physics,
  Yukawa Institute for Theoretical Physics, 
Kyoto University, Kyoto, 606-8502, Japan} 

\author{Kyohei Kawaguchi} \affiliation{Max Planck Institute for
  Gravitational Physics (Albert Einstein Institute), Am Mühlenberg 1,
  Potsdam-Golm, 14476, Germany}\affiliation{Center for Gravitational Physics,
  Yukawa Institute for Theoretical Physics, 
Kyoto University, Kyoto, 606-8502, Japan} 

\author{Koutarou Kyutoku} \affiliation{
Theory Center, Institute of Particle and Nuclear Studies, KEK,
Tsukuba 305-0801, Japan\\
Department of Particle and Nuclear Physics, the Graduate University
for Advanced Studies (Sokendai), Tsukuba 305-0801, Japan\\
Interdisciplinary Theoretical Science (iTHES) Research Group, RIKEN,
Wako, Saitama 351-0198, Japan
}\affiliation{Center for Gravitational Physics,
  Yukawa Institute for Theoretical Physics, 
Kyoto University, Kyoto, 606-8502, Japan} 

\author{Yuichiro Sekiguchi} \affiliation{Department of Physics, Toho
  University, Funabashi, Chiba 274-8510, Japan}

\author{Masaru Shibata}\affiliation{Center for Gravitational Physics,
  Yukawa Institute for Theoretical Physics, 
Kyoto University, Kyoto, 606-8502, Japan} 

\author{Keisuke Taniguchi}\affiliation{Department of Physics, University of the Ryukyus, Nishihara, Okinawa 903-0213, Japan}

\date{\today}

\newcommand{\beq}{\begin{equation}}
\newcommand{\eeq}{\end{equation}}
\newcommand{\beqn}{\begin{eqnarray}}
\newcommand{\eeqn}{\end{eqnarray}}
\newcommand{\pa}{\partial}
\newcommand{\vp}{\varphi}
\newcommand{\varep}{\varepsilon}
\newcommand{\ep}{\epsilon}
\newcommand{\comp}{(M/R)_\infty}
\begin{abstract}

Extending our previous studies, we perform high-resolution simulations
of inspiraling binary neutron stars in numerical relativity.  We
thoroughly carry through a convergence study in our currently available 
computational resources with the smallest grid spacing of $\approx
63$--86~meter for the neutron-star radius 10.9--13.7\,km.  The
estimated total error in the gravitational-wave phase is of order
0.1~rad for the total phase of $\agt 210$\,rad in the last $\sim
15$--16 inspiral orbits.  We then compare the waveforms (without
resolution extrapolation) with those calculated by the latest
effective-one-body formalism (tidal SEOBv2 model referred to as TEOB
model).  We find that for any of our models of binary neutron stars, the
waveforms calculated by the TEOB formalism agree with the numerical-relativity 
waveforms up to $\approx 3$\,ms before the peak of the
gravitational-wave amplitude is reached: For this late inspiral stage,
the total phase error is $\alt 0.1$\,rad. Although the gravitational
waveforms have an inspiral-type feature for the last $\sim 3$\,ms,
this stage cannot be well reproduced by the current TEOB formalism, in
particular, for neutron stars with large tidal deformability (i.e.,
lager radius). The reason for this is described.

\end{abstract}


\maketitle

\section{Introduction}


Gravitational-wave astronomy has vividly revealed its usefulness for
exploring the nature of compact objects. Advanced LIGO operating since
2015~\cite{aligo} has already reported three merger events of binary
black holes~\cite{abbott16,abbott16b,abbott17}, and the masses and
spins of individual black holes have been successfully determined to
modest accuracy despite the cosmological distances to these
sources~\cite{LIGOBH}. A noteworthy feature of these events,
particularly GW\,150914~\cite{abbott16}, is that they are observed
throughout the inspiral-merger-ringdown phases, which cannot be fully
modeled by the traditional post-Newtonian (PN) approximation suitable
only for the early inspiral phase~\cite{2014LRR....17....2B}. Thus,
accurate theoretical waveforms applicable to dynamical phases are
essential for the estimation of binary parameters~\cite{LIGOBH} and also
the test of general relativity~\cite{testGR}. For this purpose, the
effective-one-body (EOB) formalism calibrated by 
numerical-relativity simulations played a very important role in the
data analysis (see Ref.~\cite{LIGOBH} and references therein). As the
quality of gravitational-wave data will be further improved in the near
future with advanced Virgo~\cite{avirgo} in operation since 2017 and
upcoming KAGRA~\cite{lcgt}, accuracy of waveform models will become
more important so as to avoid systematic errors.

The next target for ground-based detectors is gravitational waves from
coalescing binary neutron stars (and also black hole-neutron star
binaries), which will inform us about finite-size properties of neutron
stars along with their masses. Simultaneous measurements of these
quantities will become a powerful method to strongly constrain the
not-yet-understood equation of state (EOS) of the neutron-star matter,
and its accomplishment is one of the most important goals of
gravitational-wave astronomy. On one hand, the masses of two neutron
stars will be determined with high accuracy of $\alt 1\%$ from
gravitational-wave signals in the inspiral phase for a sufficiently high
signal-to-noise ratio~\cite{cutler94} as far as the neutron-star spins
are small~\cite{hannam13}. On the other hand, it will be challenging to
extract quantities associated with the finite-size effect, because it
does not become appreciable until the very late inspiral phase.

Among various proposals, one of the most promising strategy is to read
off {\em tidal deformability}, which governs the late-phase orbital
evolution, from gravitational waves emitted during the inspiral phase
up to the
merger~\cite{lai94,flanagan08,hinderer10,BDF2012,Read13,wade14,aga15}. This
strategy requires an accurate template of gravitational waves from
binary-neutron-star inspirals taking into account tidal deformation
that influences the dynamics of the late inspiral orbits. In
anticipation of coming detections of binary-neutron-star mergers,
developing an accurate model of gravitational waveforms for
binary-neutron-star inspirals including the tidal effects is an urgent
task.

The EOB formalism will play an important role also in the analysis of
gravitational waves from binary neutron stars. Because the finite-size
effect becomes important in the very late stage of the orbits, PN
approximations with tidal effects~\cite{flanagan08} are not
satisfactory. Indeed, it has been shown that the lack of knowledge
about higher-order PN point-particle terms prevents us from accurately
extracting the tidal
deformability~\cite{favata14,yagi14,wade14,Hotoke15,Buonanno:1998gg,Lackey:2014fwa}. 
The EOB
formalism can improve the accuracy of the waveform templates for the
dynamical phase via effective incorporation of higher-order PN terms
and non-adiabatic dynamics close to merger. Motivated by this success,
tidal EOB (TEOB) models have been developed by various authors to
model the coalescence of binary neutron
stars~\cite{damour12,BDF2012,BD2014,bernuzzi14,hinderer16,Steinhoff:2016rfi}. These
studies demonstrate that the extraction of tidal deformability is
feasible for an event with a moderately high signal-to-noise
ratio~\cite{damour12} (this fact is also confirmed by a
numerical-relativity study~\cite{Hotoke16}), or by stacking analysis
of multiple events~\cite{aga15}. All these results support the idea
that extracting the tidal deformability from gravitational waves
emitted in the late inspiral phase of binary-neutron-star mergers is a
promising way to constrain the EOS of neutron-star matter.

However, Ref.~\cite{hinderer16} also suggests that the current TEOB
approach is not yet accurate enough to model waveforms for the last
few cycles for the case that the neutron-star radius is large. This
implies that further modeling aided by high-precision waveforms
derived by numerical relativity is required to obtain reliable
templates in the final inspiral stage of neutron star binaries. For
this purpose, a large-scale numerical-relativity simulation is
crucial.

Long-term simulations for binary-neutron-star inspirals have recently
been performed by several groups aiming at deriving high-precision
numerical-relativity 
waveforms~\cite{Read09b,baiotti11,thierfelder11,bernuzzi11,bernuzzi12,HKS2013,radice14,bernuzzi14,Hotoke15,Haas16,Hotoke16,tim17,tim17a,Bernuzzi:2015rla}. 
These work, in particular the latest ones, followed the late inspiral
phase for $\agt 10$ orbits up to the onset of merger. However, past
numerical simulations would not be able to obtain gravitational
waveforms with sufficient accuracy due to the following reasons. First
in the early-stage work, initial data with an unphysical residual
eccentricity were employed. This seriously degrades the accuracy of
derived waveforms, because binary neutron stars in the late inspiral
stage are believed to have a quasi-circular orbit with negligible
eccentricity~\cite{peters64}. This problem has been overcome, and
simulations were performed with much less eccentric initial conditions
in the latest work~\cite{Hotoke15,Haas16,Hotoke16,tim17,tim17a}.
However, even in these recent work, the phase error in the waveforms
was likely to be still of order 1\,rad because of the insufficient
grid resolutions except for a single highest resolution model of
Ref.~\cite{tim17a}.

In this paper, we push forward our previous numerical-relativity
studies~ \cite{Hotoke15,Hotoke16} to a sub-100-meter-resolution
regime. The simulations are performed for about 15--16 inspiral orbits
employing the initial data in which the eccentricity is sufficiently
small ($\sim 10^{-3}$: see appendix A) as in our previous
studies~\cite{Hotoke15,Hotoke16}. The update lies in the grid
resolution improved by a factor of up to $\sim 2.2$ from the previous
ones. In the highest-resolution case, the minimum grid spacing is
63--86\,m for the neutron stars of radius 10.9--13.7\,km: The major
diameter of neutron stars is covered by $\approx 270$ grid points. We
show that the waveform depends very weakly on the grid spacing at such
a high resolution, and the phase error in the gravitational waveforms
is estimated to be of order 0.1\,rad among the total phase of $\agt
210$\,rad. We then show that a TEOB model can be reliably calibrated
with such high-accuracy numerical gravitational waveforms.

The paper is organized as follows. In Sec.~II, we summarize the
formulation and numerical schemes employed in our numerical-relativity
study, and also review the adopted EOS. In Sec.~III, we present
numerical gravitational waveforms and show that the phase error in
gravitational waves derived with our highest grid resolution is of order
0.1\,rad. We then compare our best-resolved waveforms with those
derived by the latest TEOB approach and examine the accuracy of the TEOB
waveform in Sec.~IV. Section~V is devoted to a summary. Throughout this
paper, we employ the geometrical units of $c=G=1$ where $c$ and $G$ are
the speed of light and the gravitational constant, respectively.

\section{Summary of our setting for numerical-relativity simulation}

In this section, we summarize the formulation and numerical schemes of
our numerical-relativity simulations, EOS employed for neutron stars,
definitions of the tidal deformability for binaries, 
and our recipe for constructing a waveform. 

\subsection{Formulation, code, and models}

We follow the inspiral and early merger stages of binary neutron stars
using a numerical-relativity code, {\tt SACRA}~\cite{yamamoto08}.
Following our previous work~\cite{Hotoke15,Hotoke16}, we employ a
moving puncture version of the Baumgarte-Shapiro-Shibata-Nakamura
formalism~\cite{BSSN}, {\em locally} incorporating a Z4c-type
constraint propagation prescription~\cite{Z4c} (see
Ref.~\cite{KST2014} for our implementation) for a solution of
Einstein's equation.  In our numerical simulations, a fourth-order
finite differencing scheme in space and time is used implementing an
adaptive mesh refinement (AMR) algorithm (see Ref.~\cite{yamamoto08}
for details). In this work, we parallelized and tuned this AMR code
significantly, and this improvement enables us to perform a number of
high-resolution simulations in a relatively short time scale: As we
describe later, the grid resolution is more than twice better (i.e.,
the grid spacing is by a factor $\alt 2$ smaller) than that in our
previous work~\cite{Hotoke15}. 
The required CPU time is $540$-$650$k core hours for the highest resolution models. 

\begin{table}[t]
\centering
\caption{\label{tab1} Model name, the location of outer boundaries
  along each axis denoted by $L$, and the finest grid spacing, $\Delta
  x_{\rm finest}$, in several different grid-resolution runs. The
  model name reflects the EOS and mass of neutron stars. $\Delta
  x_{\rm finest}$ is listed for $N=182$, 150, 130, 110, 102, and 90 in
  the equal-mass models and $N=150$, 130, 110, 102, and 90 in the
  unequal-mass models.  We note that the wavelength of gravitational
  waves is initially $\lambda_0 \approx 810$\,km irrespective of the
  models in this paper. }
\begin{tabular}{lcl}
\hline\hline
~~Model~~ & ~$L$\,(km)~ & ~~~~~~~~~~~~$\Delta x_{\rm finest}$\,(m)    
\\ \hline
B135-135   & 5860 &~63,~ 76,\,\,\,\,\,  88,\,\,\, 104,\, 112,\, 127  \\
HB135-135  & 6392 &~69,~ 83,\,\,\,\,\,  96,\,\,\, 113,\, 122,\, 138  \\
H135-135   & 6991 &~75,~ 91,\,\,\,\,  105,\,    124,\, 134,\, 152  \\
125H135-135& 7324 &~79,~ 95,\,\,\,\,  110,\,    130,\, 140,\, 159  \\
15H135-135 & 7990 &~86,~ 104,\,       120,\,    142,\, 153,\, 173  \\
B121-151   & 5991 &~~~~~~~78,\,\,\,\,\, 90,\,\,\, 106,\, 114,\, 129  \\
HB121-151  & 6324 &~~~~~~~82,\,\,\,\,\, 95,\,\,\, 112,\, 121,\, 137  \\
H121-151   & 6823 &~~~~~~~89,\,\,\,\, 103,\,    121,\, 131,\, 148  \\
125H121-151& 7323 &~~~~~~~95,\,\,\,\, 110,\,    130,\, 140,\, 159  \\
15H121-151 & 7822 &~~~~~~~102,\,      118,\,    138,\, 150,\, 170  \\
\hline\hline
\end{tabular}
\end{table}

In this work, we prepare ten refinement levels for the AMR
computational domain. Specifically, two sets of four finer domains
comoving with each neutron star cover the region of their vicinity.
The other six coarser domains cover both neutron stars by a wider
domain with their origins fixed at the center of the mass of the
binary system.  Each refinement domain consists of a uniform,
vertex-centered Cartesian grid with $(2N+1,2N+1,N+1)$ grid points for
$(x,y,z)$ (the equatorial plane symmetry at $z=0$ is imposed). The
distance from the origin to outer boundaries along each axis is
denoted by $L$. Here, $L$ is much larger than the initial wavelength
of gravitational waves, $\lambda_{0} =\pi/\Omega_{0}$, with
$\Omega_{0}$ being the initial orbital angular velocity (see
Table~\ref{tab1}). We always choose it $\Omega_0 \approx 0.0155/m_0$ where
$m_0$ is the total mass of the binary system at infinite separation.

In this work, we consider the models of total mass $m_0\approx
2.7M_\odot$ (see Table~\ref{tab2}).  More precisely, we select
equal-mass models with each mass $m_1=m_2=1.35M_\odot$ and
unequal-mass models with each mass $m_1 \approx 1.21M_\odot$ and $m_2
\approx 1.51M_\odot$. For these models, the chirp mass defined by
$(m_1 m_2)^{3/5} m_0^{-1/5}$ (where $m_0=m_1+m_2$) is fixed to be
$\approx 1.17524M_\odot$. For the unequal-mass models, the symmetric
mass ratio defined by $\eta:=m_1 m_2 /m_0^2$ is chosen to be $0.247$
(i.e., the corresponding mass ratio $q=m_1/m_2$ is chosen to be
$\approx 0.8025$).  For these values of $m_0$, $\lambda_0 \approx
810$\,km and initial gravitational-wave frequency $f \approx 370$\,Hz.

The grid spacing for each domain is $\Delta x_{l}=L/(2^{l}N)$, where
$l=0-9$.  In this work, we choose a wide variety of values for $N$ and
examine the convergence properties of numerical results: For the
equal-mass models, we perform the simulations with $N=182$, 150, 130,
110, 102, and 90 and for the unequal-mass models, $N=150$, 130, 110,
102, and 90.  We note that in our previous
work~\cite{Hotoke15,Hotoke16}, $N$ was at best 72~\cite{aboutL}. 
 With the highest
grid resolution, $N=182$, the semi-major diameter of each neutron star
is covered by about 270 grid points. For the simulation with a
small-radius neutron star of radius 10.9\,km, the best grid spacing is
$\approx 63$\,m (see Table~\ref{tab1}).  With our setting of
$m_0\Omega_{0} \approx 0.0155$, the binary experiences 15--16 orbits
before the gravitational-wave amplitude reaches a peak.

\begin{table}[t]
\centering
\caption{\label{tab2} Equations of state employed, the radius,
  $R_{M}$, and the dimensionless tidal deformability $\Lambda_{M}$ of
  spherical neutron stars of $M=1.21$, 1.35, and $1.51M_\odot$.  $R_M$
  is listed in units of km.  }
\begin{tabular}{ccccccc}
\hline\hline
~EOS~ & ~$R_{1.21}$~ & ~$R_{1.35}$~ & ~$R_{1.51}$~ &
~$\Lambda_{1.21}$~ & ~$\Lambda_{1.35}$~ &~$\Lambda_{1.51}$~ 
\\ \hline
B    & 10.98 & 10.96 & 10.89 & ~581~  & ~289~  & ~131~ \\
HB   & 11.60 & 11.61 & 11.57 & ~827~  & ~422~  & ~200~ \\
H    & 12.25 & 12.27 & 12.26 & ~1163~ & ~607~  & ~298~ \\
125H & 12.93 & 12.97 & 12.98 & ~1621~ & ~863~  & ~435~ \\
15H  & 13.63 & 13.69 & 13.73 & ~2238~ & ~1211~ & ~625~ \\
\hline\hline
\end{tabular}
\end{table}

We prepare binary neutron stars in quasi-circular orbits with small
eccentricity $\sim 10^{-3}$ for the initial condition of numerical
simulations. These initial conditions are numerically obtained by
using a spectral-method library, LORENE~\cite{lorene}.  The
eccentricity reduction is performed by the method of
Ref.~\cite{KST2014}.  The neutron stars are assumed to have an
irrotational velocity field, which is believed to be an
astrophysically realistic (or at least approximately realistic) 
configuration~\cite{bildsten92,kochaneck92}. 

We note that even with the eccentricity-reduced initial conditions,
the small residual eccentricity of $\sim 10^{-3}$ still
gives a small damage for getting accurate quasi-circular waveform.
This is in particular the case for carefully comparing the numerical
waveforms with those by TEOB formalisms (see the discussion in
Appendix A). The numerical waveforms for the first 3--4 orbits are not
very suitable for performing the careful analysis of
gravitational-wave data. Thus, when comparing the numerical waveforms
with those by the TEOB formalisms, we only employ the waveforms for
the last 11--12 orbits.  We first notice this fact when we obtain the
results of high-resolution simulations in this paper. This finding
reconfirms that the eccentricity reduction for constructing the
initial data is crucial for accurately deriving the late inspiral 
waveforms.

\subsection{Equations of State}

Following our previous work~\cite{Read09b,lackey12,Read13}, we employ
a parameterized piecewise-polytropic EOS~\cite{read09} with two
pieces. In this work, our purpose is to accurately clarify the
dependence of inspiral gravitational waveforms on the tidal
deformability. For this purpose, the choice of the simple EOS is
acceptable.

This EOS is written in terms of two segments of polytropes of
the form
\begin{align}
P = \Bigg\{
\begin{array}{l}
K_{0}\rho^{\Gamma_{0}} ~~~ \text{( for $\rho_{0}\leq \rho<\rho_{1}$)} \\
K_{1}\rho^{\Gamma_{1}} ~~~ \text{( for $\rho_{1}\leq \rho$)} \\
\end{array}
\end{align}
where $\rho$ is the rest-mass density, $P$ is the pressure, $K_{0}$ and $K_{1}$ are  
a polytropic constant, and $\Gamma_{0}$ and $\Gamma_{1}$ are an adiabatic index.  At the
boundary of these two piecewise polytropes, $\rho=\rho_{1}$, the
pressure is required to be continuous, i.e.,
$K_{0}\rho_{1}^{\Gamma_{0}}=K_{1}\rho_{1}^{\Gamma_{1}}$. Thus, the
parameters, which have to be given, are $K_0$, $\rho_1$, $\Gamma_0$,
and $\Gamma_1$.  Following the previous
studies~\cite{Read09b,lackey12,Read13}, these parameters are
determined in the following manner: The low-density EOS is fixed by setting
$\Gamma_{0}=1.3562395$ and $K_{0}=3.594\times 10^{13}$ in cgs
units. The adiabatic index for the high-density region is set to be
$\Gamma_1=3$, and hence, $K_1$ is determined to be
$K_1=K_0\rho_1^{\Gamma_0-\Gamma_1}$.  The remaining parameter,
$\rho_1$, is varied for a wide range to prepare neutron stars with a
variety of the radius and tidal deformability (see Table~\ref{tab2}).
We note that for any EOS employed in this paper, the maximum mass of
spherical neutron stars is larger than $2.0M_\odot$; the approximate
maximum mass for neutron stars for which the mass is accurately
measured to date~\cite{demorest10} (see Ref.~\cite{lackey12} on the
data of the maximum mass for each EOS).

In numerical simulations, we employ the following modified version of the
piecewise polytropic EOS to approximately take into account thermal
effects:
\beqn
P &=& P_{\rm cold}(\rho) +( \Gamma_{\rm th} - 1 ) \rho \varepsilon_{\rm th}, \\
\varepsilon &=& \varepsilon_{\rm cold}(\rho) + \varepsilon_{\rm th} ,
\eeqn
where $\Gamma_{\rm th}$ is a constant.  The cold parts (the first terms) of
both variables are calculated using the original piecewise polytropic
EOS from $\rho$, and then the thermal part of the specific internal
energy is determined from $\varepsilon$ as $\varepsilon_{\rm th} =
\varepsilon - \varepsilon_{\rm cold}(\rho)$. Because $\varepsilon_{\rm
  th}$ vanishes in the absence of shock heating, it is regarded as the
finite-temperature part determined by the shock heating in the present
context.  Following our latest work~\cite{KST2014,Hotoke15,Hotoke16},
$\Gamma_{\rm th}$ is chosen to be 1.8, but this is not relevant for
the present study because we focus only on the late inspiral
evolution.

\subsection{Parameters associated with tidal deformability}

For modeling the late inspiral orbital motion and corresponding
gravitational waves of binary neutron stars, two parameters
constructed from the dimensionless tidal deformability of two neutron
stars are often used. One is the EOB tidal parameter~\cite{damour12}
which appears in the equation of motion of the TEOB formalism and is
defined by
\beqn
\Lambda_T:=16 \eta (X_1^3 \Lambda_1 + X_2^3 \Lambda_2),
\eeqn
where $X_i:=m_i/m_0$ and $\Lambda_i$ is the dimensionless tidal
deformability of each neutron star.  We note that the originally
defined variable is $\kappa^T_2$ and it is calculated by
$3\Lambda_T/16$.

The other parameter is the so-called binary tidal deformability,
$\tilde \Lambda$, defined by~\cite{wade14}
\beqn
\tilde \Lambda={8 \over 13} \biggl[
 && (1 + 7\eta -31 \eta^2)(\Lambda_1 + \Lambda_2)  \nonumber \\
&&  -\sqrt{1-4\eta}(1+9\eta-11\eta^2)(\Lambda_1-\Lambda_2) \biggr].  
\eeqn
where we supposed that $m_1 \le m_2$ and then, $\Lambda_1 \ge \Lambda_2$.
$\tilde\Lambda$ is related to the leading term associated with the
tidal effect in the gravitational-wave phase in the Fourier space, and
hence, in the real gravitational-wave detection, this can be regarded
as the primarily measured quantity.

One interesting property for $\Lambda_T$ and $\tilde \Lambda$ is that
for a fixed chirp mass, these values are in a narrow range (within 1\%
disagreement) as long as we consider the cases that $m_0 \approx
2.7M_\odot$ and $0.8 \leq q \leq 1$.  Hence, in the following, we
refer only to $\tilde \Lambda$, supposing that $\tilde \Lambda$ agrees
approximately with $\Lambda_T$. Note that for the equal-mass models 
$(\eta=0.25)$, $\Lambda_T=\tilde{\Lambda}$. 

\begin{figure*}[t]
\begin{center}
\includegraphics[width=84mm]{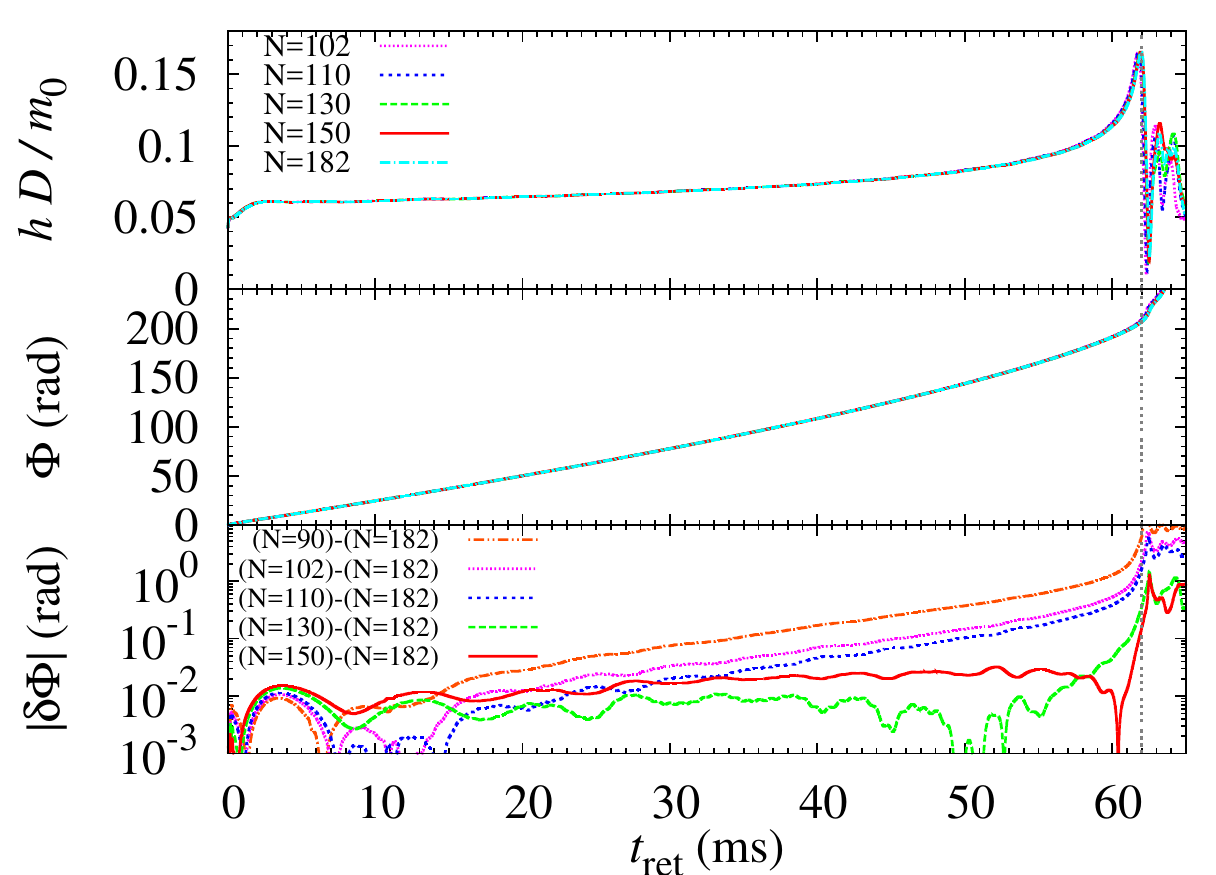}~~~
\includegraphics[width=84mm]{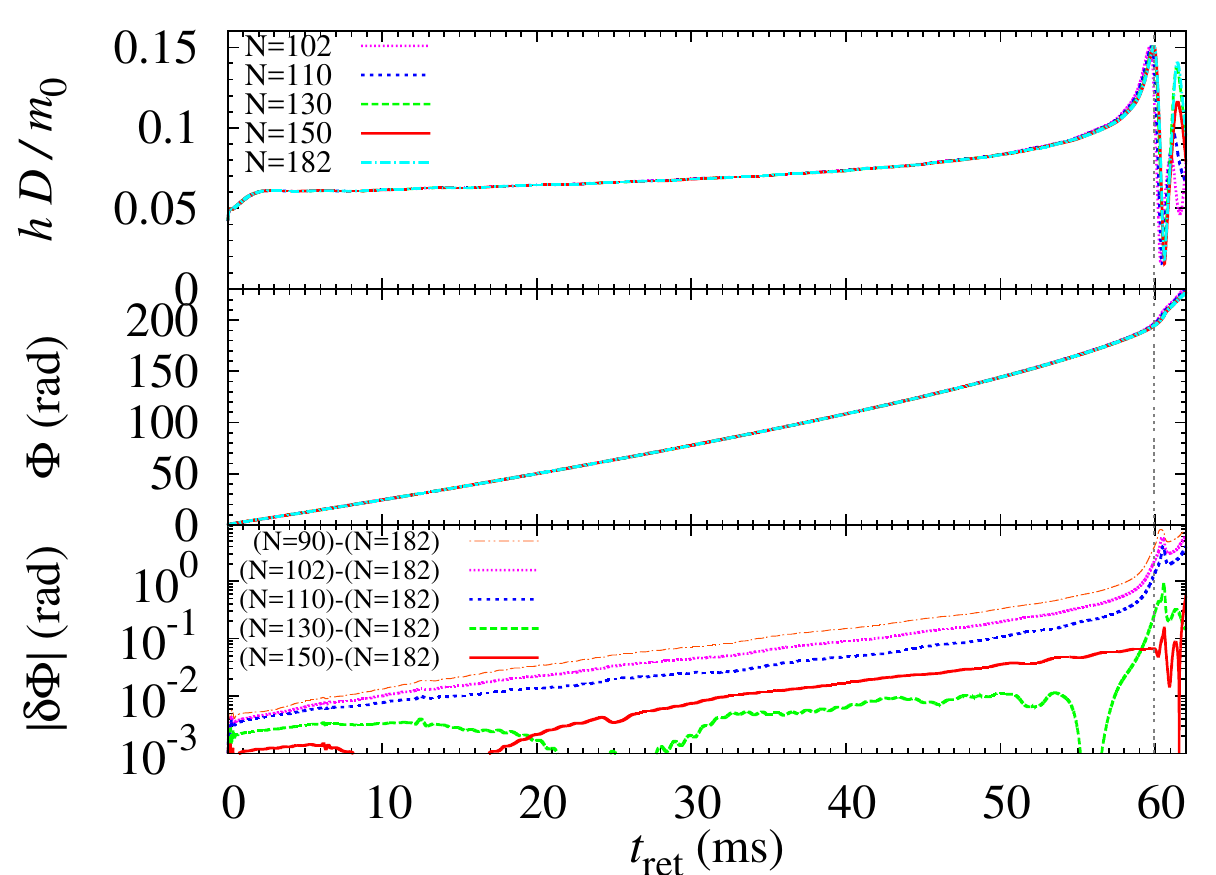}
\caption{The amplitude ($A^{2,2}D/m_0$; upper panels) and phase
  ($\Phi$; middle panels) of numerical gravitational waveforms with
  different grid resolutions for the equal-mass models with HB EOS
  (left) and 125H EOS (right).  In the bottom panels, we also plot the
  difference in phase with respect to the best grid resolution
  ($N=182$ case) for $N=90$, 102, 110, 130, and 150.  The vertical
  lines denote the time at which the peak amplitude is reached for
  $N=182$. $D$ denotes the distance to the source. 
\label{fig1}}
\end{center}
\end{figure*}

\subsection{Extraction of gravitational waves}

As a first step toward producing gravitational waves from numerical
data, we extract the outgoing-component of complex Weyl scalar
$\Psi_{4}$~\cite{yamamoto08}.  If $\Psi_4$ is extracted at a
sufficiently large radius, complex gravitational waveforms are
determined in spherical coordinates $(r , \theta , \phi )$ by
\begin{eqnarray}
  h(t,\theta,\varphi) = -\lim_{r\rightarrow \infty}
\int^{t}dt^{\prime}\int^{t^{\prime}}dt^{\prime \prime}
\Psi_{4}(t^{\prime \prime},r,\theta,\varphi).~~\label{eq:hpsi4}
\end{eqnarray}
$\Psi_{4}$ can be expanded with respect to the spin-weighted spherical
harmonics of weight $-2$, $_{-2}Y_{lm}$, as
\begin{eqnarray}
  \Psi_{4}(t, r, \theta, \phi) = \sum_{lm}\Psi_{4}^{l,m}(t,r)
  {}_{-2}Y_{lm}(\theta, \phi),
\end{eqnarray}
where $\Psi_{4}^{l,m}$ is the expansion coefficient defined
by this equation.  In this work, we focus only on the $(l,|m|)=(2,2)$
mode because we pay attention only to the equal-mass or nearly
equal-mass binary, and hence, this quadrupole mode is the dominant
one. 

We extract $\Psi_{4}$ at a finite spherical-coordinate radius of $r
\approx 200m_0$, and then, calculate $\Psi_4^{2,2}$ as a function of
the retarded time defined by
\begin{eqnarray}
t_{\rm{ret}} := t - r_{*}, \label{eq:tret1}
\end{eqnarray}
where $r_{*}$ is the so-called tortoise coordinate defined by
\begin{eqnarray}
r_{*} := r_{\rm{A}} + 2m_0\ln \left(\frac{r_{\rm{A}}}{2m_0}-1\right),
 \label{eq:tret2}
\end{eqnarray}
with $r_{\rm{A}}:=\sqrt{A/4\pi}$ and $A$ the proper area of the
extraction sphere.

Since $\Psi_4^{2,2}(t_{\rm ret})$ extracted at a finite radius,
$r_0 \approx 200m_0$, 
is different from the true gravitational waveform observed at null
infinity, we then have to compute an extrapolated waveform at $r_0
\rightarrow \infty$. As in our previous
studies~\cite{Hotoke15,Hotoke16}, for this purpose, we employ the
Nakano's method (approximately equivalent to the Cauchy matching
method)~\cite{LNZC2010,Nakano15,Nakano:2015pta}, by which the waveform at infinity is
calculated by
\beqn
\Psi_4^{l,m,\infty}(t_{\rm ret}, r_0)
&=&C(r_0)\left[\Psi_4^{l,m}(t_{\rm ret}, r_0) \right.\nonumber \\
&~&\left. -{(l-1)(l+2) \over 2r_{\rm A}}
\int^{t_{\rm ret}} \Psi_4^{l,m}(t', r_0)dt'
\right], \nonumber \\
\label{eq:nakano}
\eeqn
where $C(r_0)$ is a function of $r_0$.  Since our coordinates are
similar to isotropic coordinates of non-rotating black holes, we
choose $r_{\rm A}=r_0[1+m_0/(2r_0)]^2$ and $C(r_0)=1 - 2m_0/r_{\rm
  A}$. In this setting, $t_{\rm ret}$ at $r=r_0$ is given by
Eqs.~(\ref{eq:tret1}) and (\ref{eq:tret2}).

We also perform the same analysis choosing different extraction radii
as $r_0/m_0=156$ and 178 and estimate the error of the
gravitational-wave phase coming from the extraction of $\Psi_4$ at
finite radii, because the effect of the finite-radius extraction still
remains in $\Psi_4^{l,m,\infty}$. 

For $\Psi_4^{l,m,\infty}(t_{\rm ret}, r_0)$ thus determined, the
gravitational waveform of each mode is obtained by twice integrating
it as [see Eq.~(\ref{eq:hpsi4})]
\beqn 
h^{l,m}(t_{\rm ret})&:=&h_+^{l,m}(t_{\rm ret}) - i h_\times^{l,m}(t_{\rm ret})
\nonumber \\
&=&-\int_{t_{\rm ret}} dt' \int_{t'} dt'' \Psi_4^{l,m,\infty}(t'').  
\eeqn
For this integration, we employ the method of Ref.~\cite{RP2011},
and write $h^{l,m}(t_{\rm ret})$ as
\beqn 
h^{l,m}(t_{\rm ret})=\int d\omega' {\Psi_4^{l,m,\infty}(\omega')
  \over {\rm max}(\omega', \omega_{\rm cut})^2} \exp(i \omega' t_{\rm
  ret}), 
\label{eq:h}
\eeqn
where $\Psi_4^{l,m,\infty}(\omega)$ is the Fourier transform of
$\Psi_4^{l,m,\infty}(t_{\rm ret})$ and $\omega_{\rm cut}$ is chosen to
be $1.6\Omega_0$. (Note that at the initial stage, the value of
$\omega$ is $\approx 2 \Omega_0 > \omega_\mathrm{cut}$).  We recall
again that in this paper we pay attention only to $l=|m|=2$ modes
because these are the dominant modes in particular for the equal-mass
binaries.

From Eq.~(\ref{eq:h}), the evolution of the amplitude, i.e.,
$A^{l,m}=|h^{l,m}|$, is immediately determined.  For the analysis
employed in our method, we can also define the angular
frequency
\beqn
\omega(t_{\rm ret}):={|\dot h^{2,2}| \over |h^{2,2}|}, \label{eq:omega}
\eeqn
and subsequently, the gravitational-wave phase by
\beqn
\Phi(t_{\rm ret}):=\int^{t_{\rm ret}} dt' \, \omega(t'). 
\eeqn
Now, using $A^{2,2}$ and $\Phi$, the quadrupole gravitational waveform 
can be written as
\beqn
h^{2,2}(t_{\rm ret})=A^{2,2}(t_{\rm ret})\exp\left[i\Phi(t_{\rm
    ret})\right].
\eeqn

Before closing this section, we note that the
angular frequency defined by Eq.~\eqref{eq:omega} is contaminated by the
time derivative of the amplitude. Although the associated error 
for our current analysis that focuses only on the inspiral phase is negligible,
future analysis of more dynamical merger phase will require more
appropriate definitions such as~\cite{kawaguchi17}
\beqn
\omega_{\rm dyn} (t_{\rm ret}):={\rm Im}\left({h^{*2,2} \dot h^{2,2} \over
|h^{2,2}|^2}\right),
\eeqn
where $h^{*2,2}$ is the complex conjugate of $h^{2,2}$.

\section{Numerical results}

\begin{figure*}[th]
\begin{center}
\includegraphics[width=84mm]{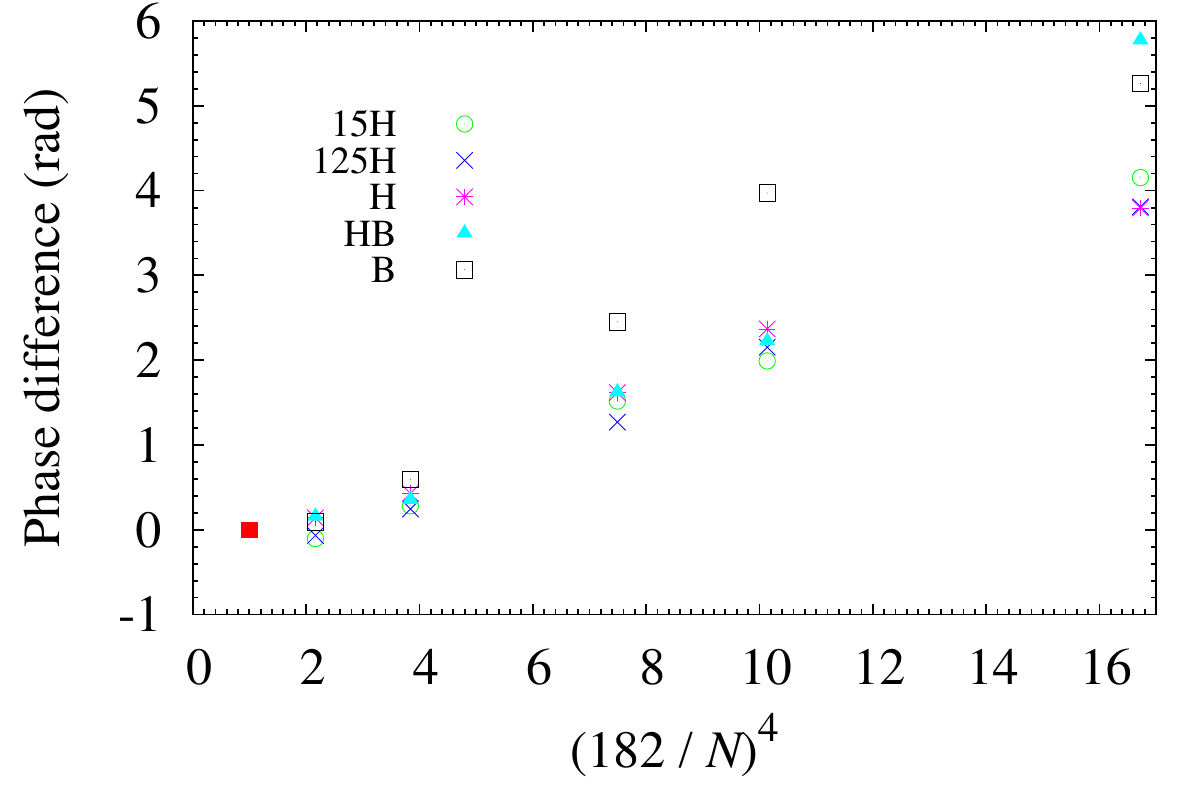}~~~
\includegraphics[width=84mm]{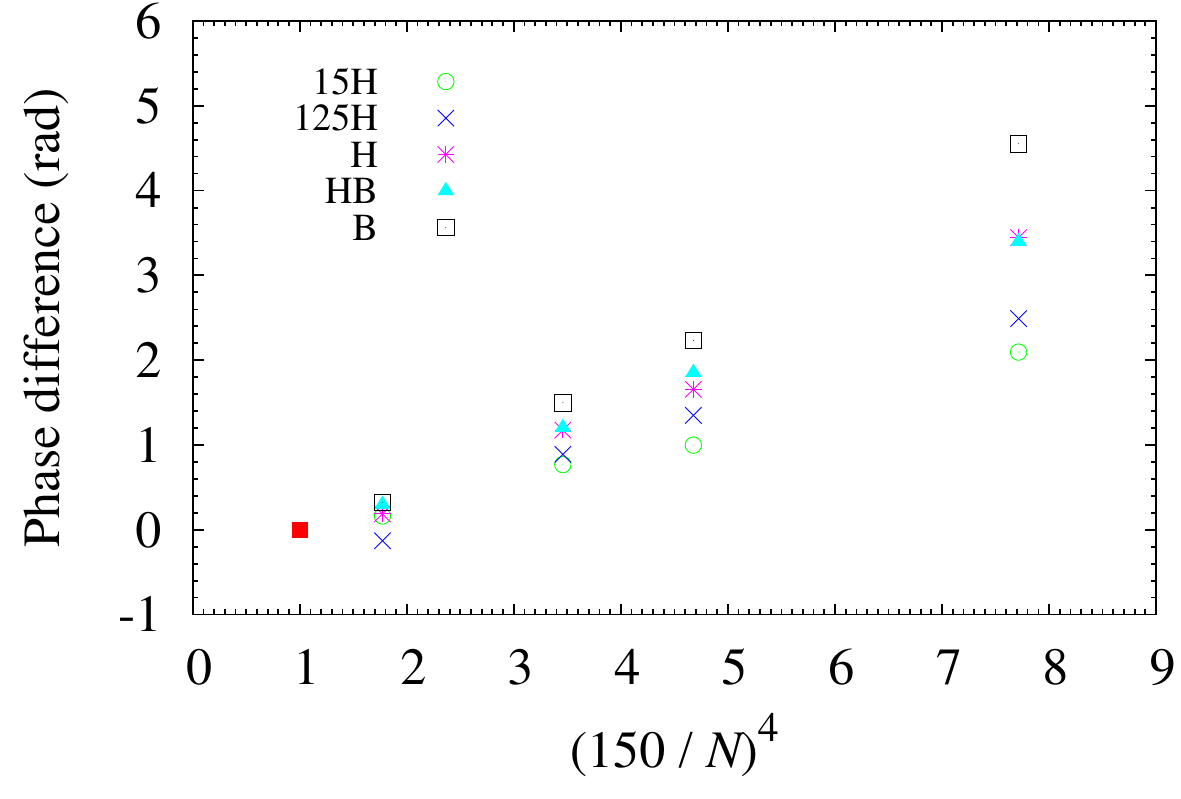}
\vspace{-4mm}
\caption{Difference in the gravitational-wave phase, $\Phi(t)$,
  between the results of the best-grid resolution and others at the
  moment that the gravitational-wave amplitude for the best-grid
  resolution reaches the peak as a function of $(182/N)^4$ for the
  equal-mass models (left) and as a function of $(150/N)^4$ for the
  unequal-mass models (right) with 5 different EOS models.
\label{fig2}}
\end{center}
\end{figure*}

\begin{figure*}[th]
\begin{center}
\includegraphics[width=84mm]{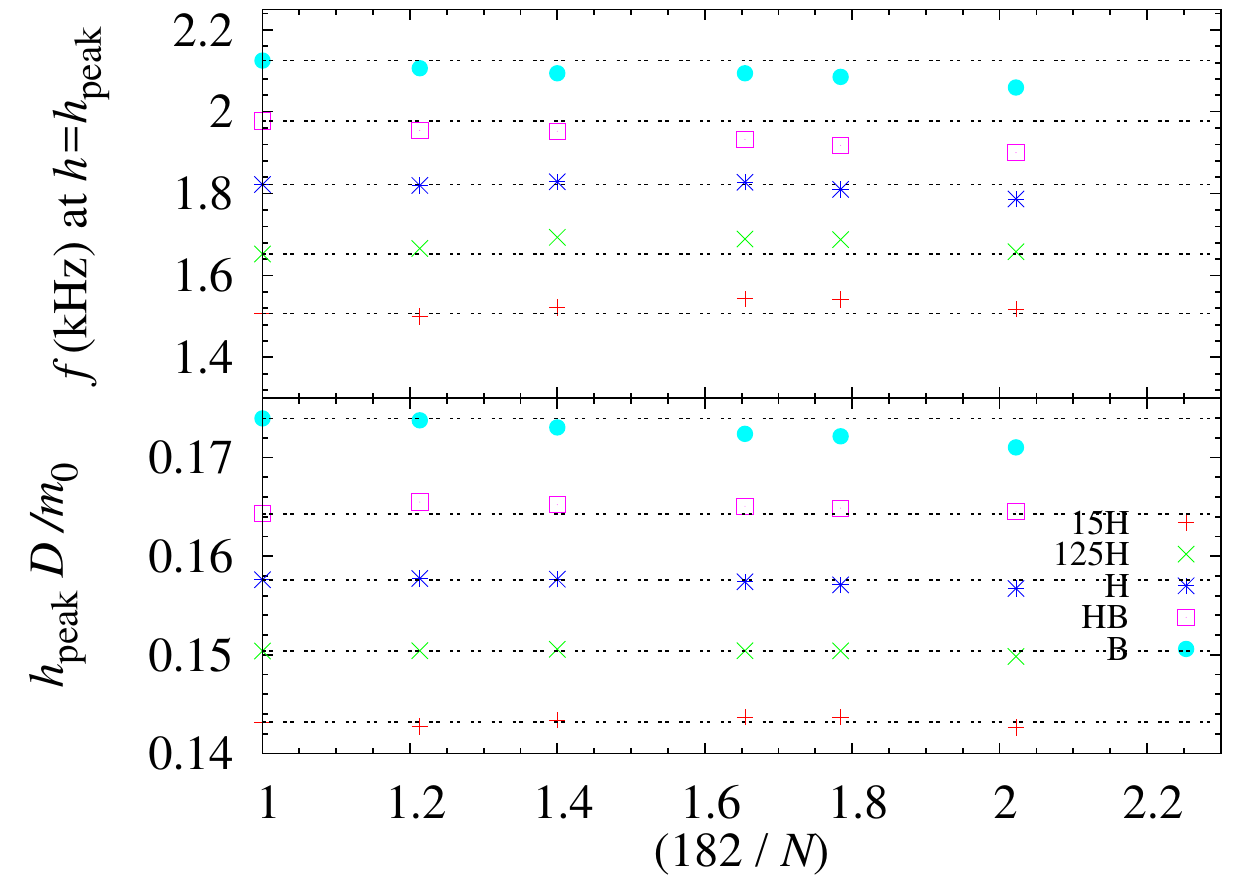}~~~
\includegraphics[width=84mm]{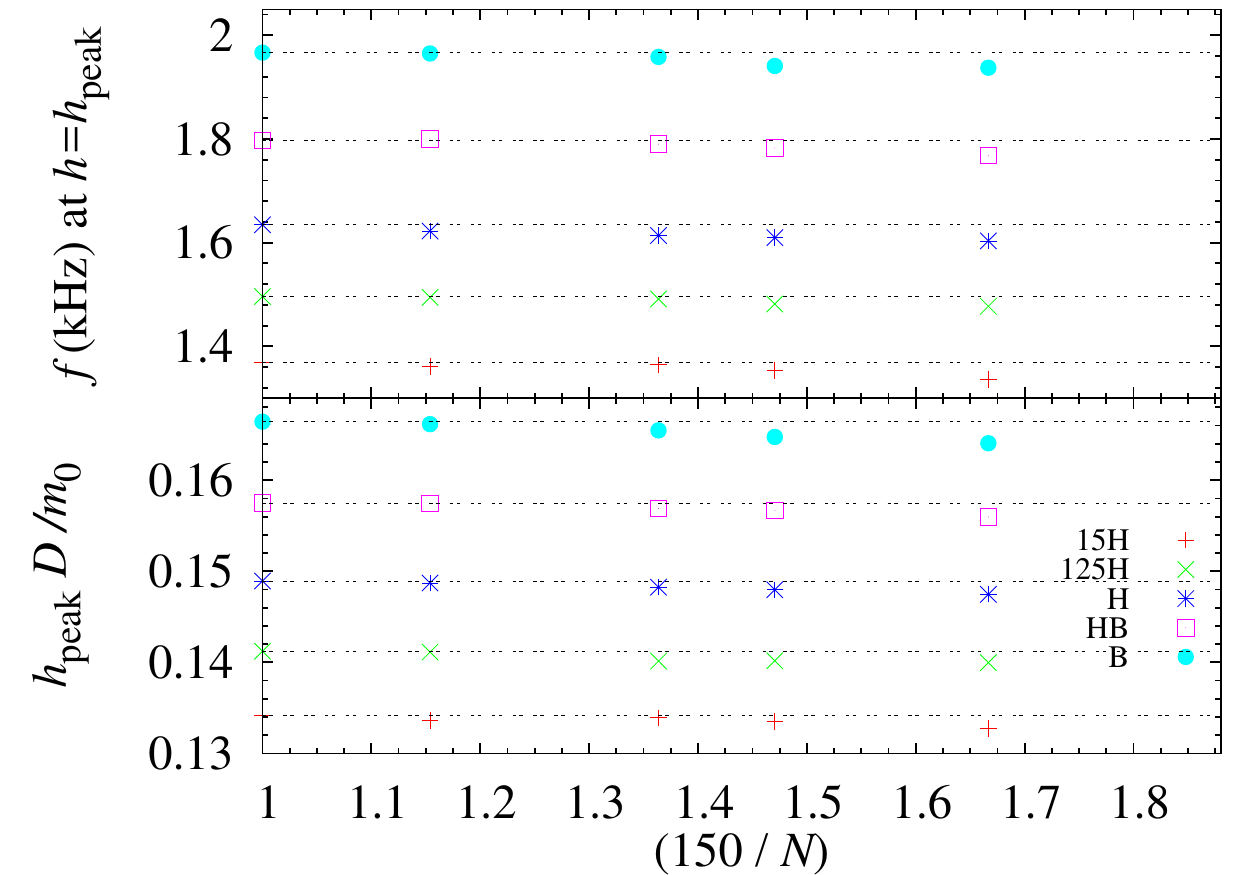}
\caption{The frequency of gravitational waves at the peak (top panel) and 
  peak amplitude of gravitational waves (bottom panel) as
  functions of $182/N$ for the equal-mass models (left panel) and as
  functions of $150/N$ for the unequal-mass models (right
  panel). The dot-dot horizontal lines denote the values of the best
  grid resolution models.
\label{fig3}}
\end{center}
\end{figure*}

Figure~\ref{fig1} plots the amplitude ($A^{2,2}D/m_0$; upper panels)
and phase ($\Phi$; middle panels) of numerical gravitational waveforms
with different grid resolutions for the equal-mass models with HB
EOS (left) and 125H EOS (right). 
$D$ denotes the distance to the source. 
In the bottom panels for both left
and right, we also plot the difference in phase with respect to the
best grid-resolution ($N=182$) results for $N=90$, 102, 110, 130, and
150. This figure shows that the merger occurs earlier for the poor
grid resolutions with $N \alt 130$.  Specifically, the evolution of
$\Phi(t_{\rm ret})$ is spuriously accelerated for such low grid
resolutions.  However, for the high grid resolutions with $N \agt
150$, the phase evolution depends only weakly on the grid
resolution. In some models like 15H and 125H models, the merger for
$N=182$ occurs slightly earlier than for $N=150$. 
However, the peak amplitude time difference is as small as $\approx 0.5$ microsecond. 
When we pay
attention to the waveforms only up to the peak amplitude, the phase
difference between $N=150$ and $N=182$ is 0.1--0.2\,rad irrespective
of the models (see Fig.~\ref{fig2}).

\begin{figure}[t]
\begin{center}
\includegraphics[width=84mm]{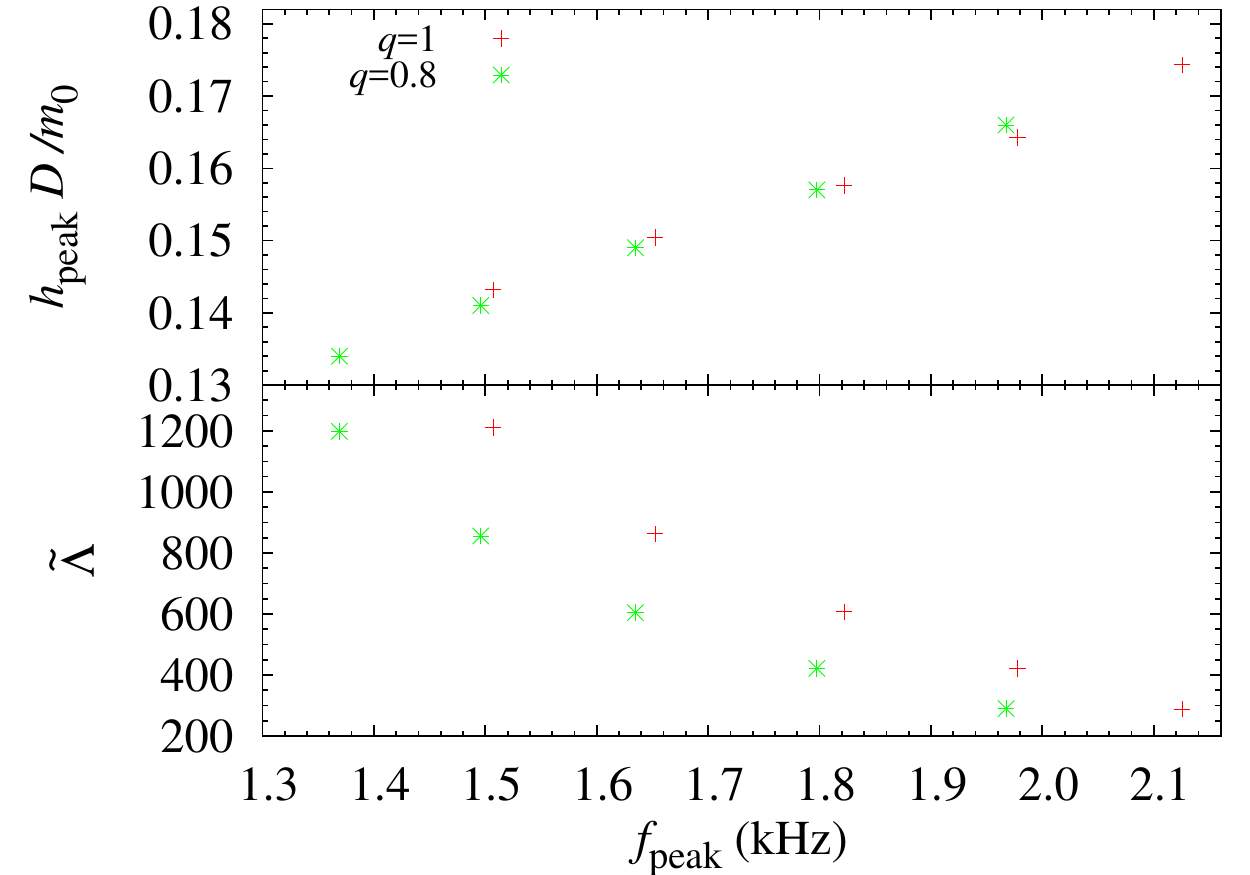}
\caption{Relation between the peak amplitude and the corresponding
  frequency of gravitational waves (top panel), and between the value
  of $\tilde \Lambda$ and the peak frequency (bottom panel). 
  We show the result of $N=182$ for the equal-mass models and $N=150$ for the unequal-mass models.
\label{fig4}}
\end{center}
\end{figure}

The left panel of Fig.~\ref{fig2} plots differences in the
gravitational-wave phase, $\Phi(t_{\rm ret})$, at the moment that the
gravitational-wave amplitude for $N=182$ reaches the peak as a
function of $(182/N)^4$ for the equal-mass models with five different
EOS.  This shows that the phase difference steeply (as fast as or
faster than the fourth-order convergence) decreases with the
improvement of the grid resolution for $N \agt 100$ irrespective of
the EOS.  In particular for $N \agt 150$, the phase difference is
decreased to 0.1--0.2\,rad, and the phase error appears to be
convergent (besides an irregular error that does not converge
monotonically with the improvement of the grid resolution: see below
for a discussion). This indicates that for $N \agt 150$ $(\Delta x_9 \lesssim 100\,{\rm m})$, nearly
convergent waveforms with the phase error within $\sim 0.2$\,rad
would be obtained.

The right panel of Fig.~\ref{fig2} plots differences in the
gravitational-wave phase for the unequal-mass models.  It is found
that the convergence behavior of the phase difference is similar to
that for the equal-mass models.  This suggests that for $N=150$, a
nearly convergent waveform is likely to be also obtained for these
unequal-mass models. 

Figure~\ref{fig1} shows that for $N \alt 110$, the phase difference
monotonically decreases with the improvement of the grid
resolution. However, for $N \agt 130$, the phase difference does not
show such monotonic behavior as already mentioned. This indicates the
presence of an unidentified source of the numerical error that does not
monotonically converge with the improvement of the grid
resolution. Figure~\ref{fig1} indicates that such error source
generates the error in gravitational-wave phase by $\sim 0.1$\,rad
irrespective of the EOS employed. 

The another phase error sources are the finite-radius extraction and 
the violation of the baryon mass conservation. 
In Appendix B, we show the phase error due to the finite-radius 
extraction is less than 0.04\,rad irrespective of the models. 

The violation of the baryon mass conservation may cause a phase error~\cite{tim17}. 
We also show that this error is much smaller than 0.01\,rad up to the merger irrespective of the models 
in the appendix. Therefore, we have to bear in mind
that in our current numerical waveform, the phase error of $\sim
0.1$\,rad cannot be avoided.

Figure~\ref{fig3} plots the peak amplitude of gravitational waves,
$h_{\rm peak}$, and the frequency at the peak, $f_{\rm peak}$, as
functions of the grid resolution, described by $(182/N)$, for the
equal-mass models (left panel) and as functions of $(150/N)$ for the
unequal-mass models (right panel). This figure shows that the
quantities associated with the peak depend weakly on the grid
resolution.  In particular, it is found that $f_{\rm peak}$ may be
underestimated if the grid resolution is not high enough. It should be
also remarked that the fluctuation in $f_{\rm peak}$ is rather large
even among the high-resolution results.  This is reasonable because
the frequency rapidly increases near the amplitude peak.  We should
keep in mind that the value of the peak frequency has an error of
2--3\%.

Figure~\ref{fig4} plots the relation between $h_{\rm peak}$ and
$f_{\rm peak}$ and between $\tilde \Lambda$ and $f_{\rm peak}$
following Ref.~\cite{Read13}. This shows that the relation between
$h_{\rm peak}$ and $f_{\rm peak}$ depends only weakly on the mass
ratio. Qualitatively this is reasonable because $h_{\rm peak}$ should
be an increasing function of $f_{\rm peak}$. It is also found that the
relation between $\tilde \Lambda$ and $f_{\rm peak}$ depends strongly
on the mass ratio.  We note that $\tilde \Lambda$ is approximately
equal to $\Lambda_T$ for the models employed in this paper. Thus, our
results do not show a universal relation (see also
Ref.~\cite{RT2016}).  Nevertheless, Fig.~\ref{fig4} shows that $f_{\rm
  peak}$ (and hence $h_{\rm peak}$) has valuable information on
$\tilde \Lambda$ as follows: (i) if $f_{\rm peak}$ is higher than
$\sim 2$\,kHz (i.e., $h_{\rm peak}D/m_0 \agt 0.16$) for the chirp mass
of $\approx 1.1752M_\odot$, $\tilde \Lambda \alt 500$, implying that
the EOS is rather soft.  (ii) if $f_{\rm peak}$ is lower than $\sim
1.4$\,kHz (i.e., $h_{\rm peak}D/m_0 \alt 0.14$) for the chirp mass of
$\approx 1.1752M_\odot$, $\tilde \Lambda \agt 1000$, implying that the
EOS is rather stiff.

\section{Comparison between numerical-relativity and TEOB
 waveforms}

\begin{figure*}[t]
\begin{center}
  \includegraphics[width=84mm]{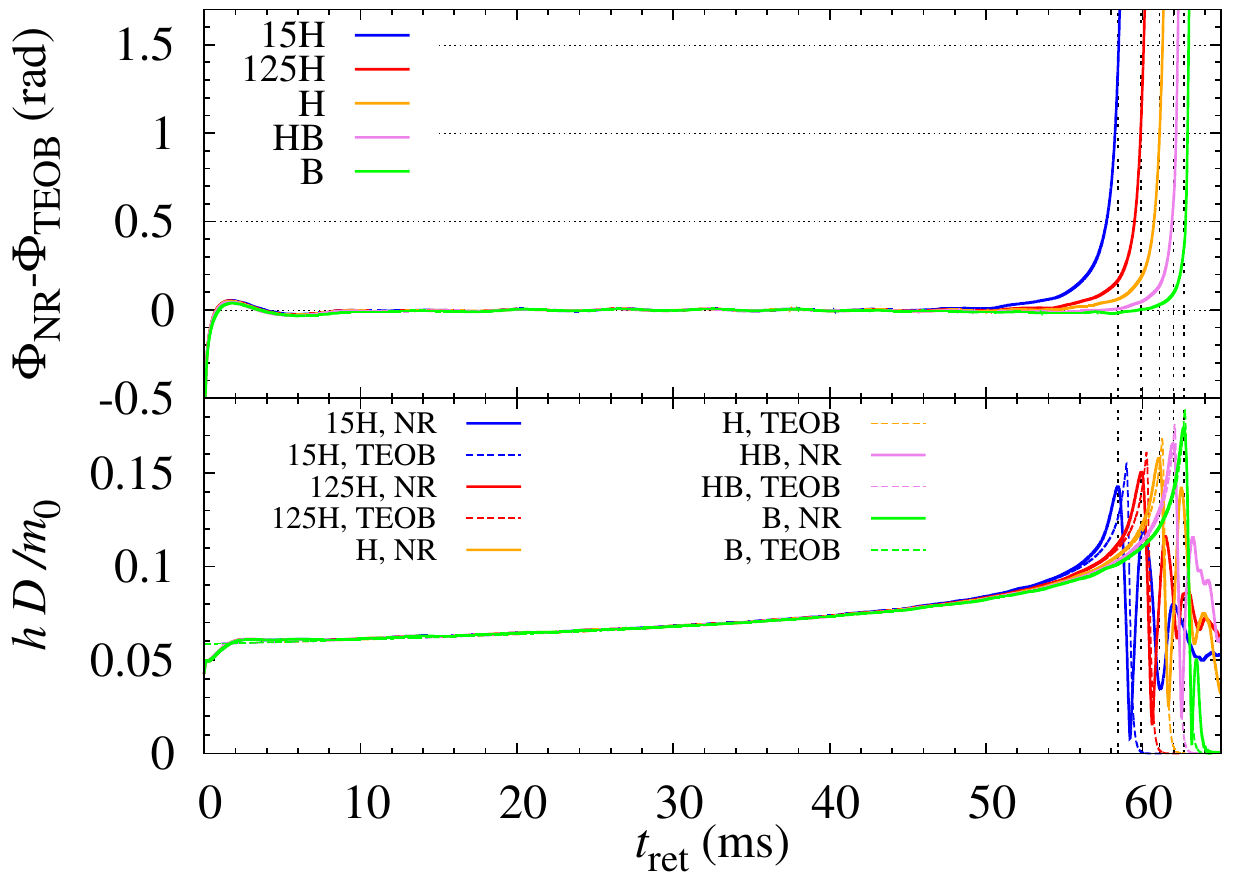}~~~
  \includegraphics[width=84mm]{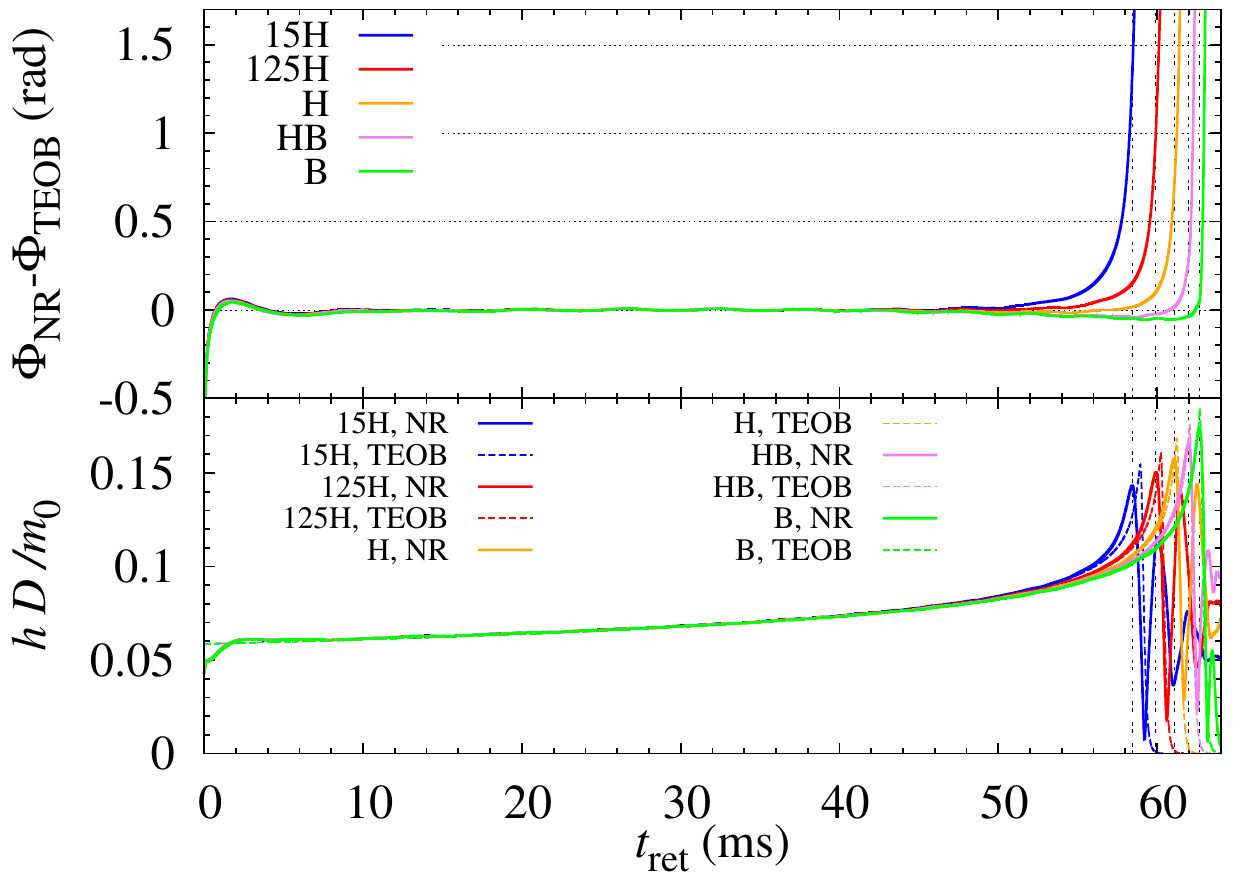}
\caption{Comparison between NR and TEOB waveforms for the equal-mass models. For the
  NR waveforms, we use the data with $N=150$ (left) and 182
  (right).  The top and bottom panels show the phase difference,
  $\Phi_{\rm NR}-\Phi_{\rm TEOB}$, and amplitude, $A^{2,2}D/m_0$ of
  gravitational waves, respectively. Here, $D$ denotes the distance to the source. 
 The vertical dashed lines are the peak amplitude time $t_{\rm peak}$ when the gravitational-wave amplitude reaches a peak in each model.
\label{fig5}}
\end{center}
\end{figure*}

\begin{figure}[th]
\begin{center}
\includegraphics[width=84mm]{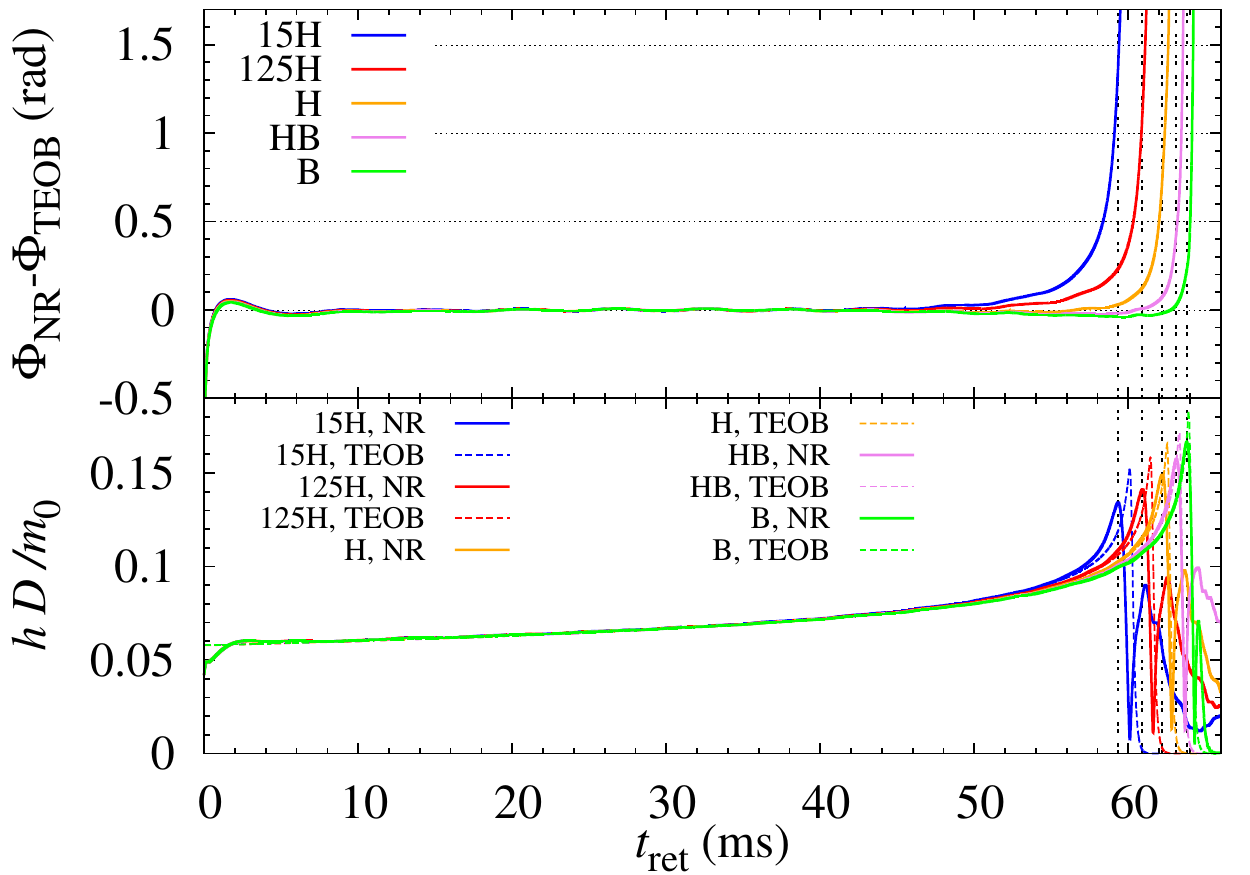}
\caption{The same as Fig.~\ref{fig5} but for the unequal-mass models. 
  For the NR waveforms, we use the data with $N=150$. 
\label{fig6}}
\end{center}
\end{figure}

In this section, we compare the NR waveforms with
those by a TEOB formalism.  For the TEOB formalism, we employ a latest
version reported in Ref.~\cite{hinderer16}, in which the effects of
not only the static but also dynamical tides are taken into account.

In this comparison, we use the NR waveforms without performing
extrapolation to the limit of $\Delta x_{\rm finest}=0$ because as we
showed in the previous section (and found in this section), the
numerical error of the waveforms is small enough to perform the direct
comparison. In the following, we employ the NR waveforms
obtained for $N=182$ or $N=150$ for the comparison: For the equal-mass
models, we use the two waveforms of different grid resolutions and
show that the results on the comparison do not lead to any serious
difference.

When comparing two waveforms, we first have to align the time and
phase of the NR and TEOB waveforms.  This is done by searching
for the minimum of the following correlation, $I_c$, varying $\tau$
and $\phi$:
\beqn
I_c&=&\min_{\tau,\phi}\int_{t_i}^{t_f} dt_{\rm ret} 
\left|A^{2,2}_{\rm NR}(t_{\rm ret}) 
\exp\left[i \Phi_{\rm NR}(t_{\rm ret})\right] \right. \nonumber \\
&&~  - \left. 
A_{\rm TEOB}^{2,2}(t_{\rm ret}+\tau)
\exp\left[i \Phi_{\rm TEOB}(t_{\rm ret}+\tau)+i \phi\right]
\right|^2. \nonumber \\
\label{corre2}
\eeqn
Here, $A^{2,2}_{\rm NR}$ and $\Phi_{\rm NR}$ denote the amplitude and
phase of gravitational waves for the numerical-relativity data, respectively. 
$A^{2,2}_{\rm TEOB}$ and $\Phi_{\rm TEOB}$ denote those by the TEOB
formalism.  For calculating the correlation, we employ the time domain 
NR waveforms of $20\,{\rm ms} \leq t_{\rm ret} \leq
40\,{\rm ms}$. The corresponding gravitational-wave frequency at 
$t_{\rm ret}=20$ and 40\,ms is $\approx 410$ and 500\,Hz, respectively. 

The reason that we choose the rather late-time NR waveforms of
$20\,{\rm ms} \leq t_{\rm ret} \leq 40\,{\rm ms}$ for the correlation
$I_c$ is as follows: In the early stage of the numerical evolution
with $t_{\rm ret} \alt 15$\,ms, the frequency of gravitational waves
always has an irregular modulation (see Appendix A). For precisely
comparing the NR waveforms with those by the TEOB approach,
such modulation, even if its amplitude is not very large, introduces
the uncertainty in matching. To remove such uncertainty, we
discard the waveforms in the early stage. We note that even for
$t_{\rm ret} \geq 20$\,ms, there are $\agt 22$ wave cycles ($\agt 11$
orbits) in our numerical data. 
We confirmed that the choice of the time-window of the matching does not significantly affect the following result. 

\begin{figure*}[t]
\begin{center}
  \includegraphics[width=50mm,angle=270]{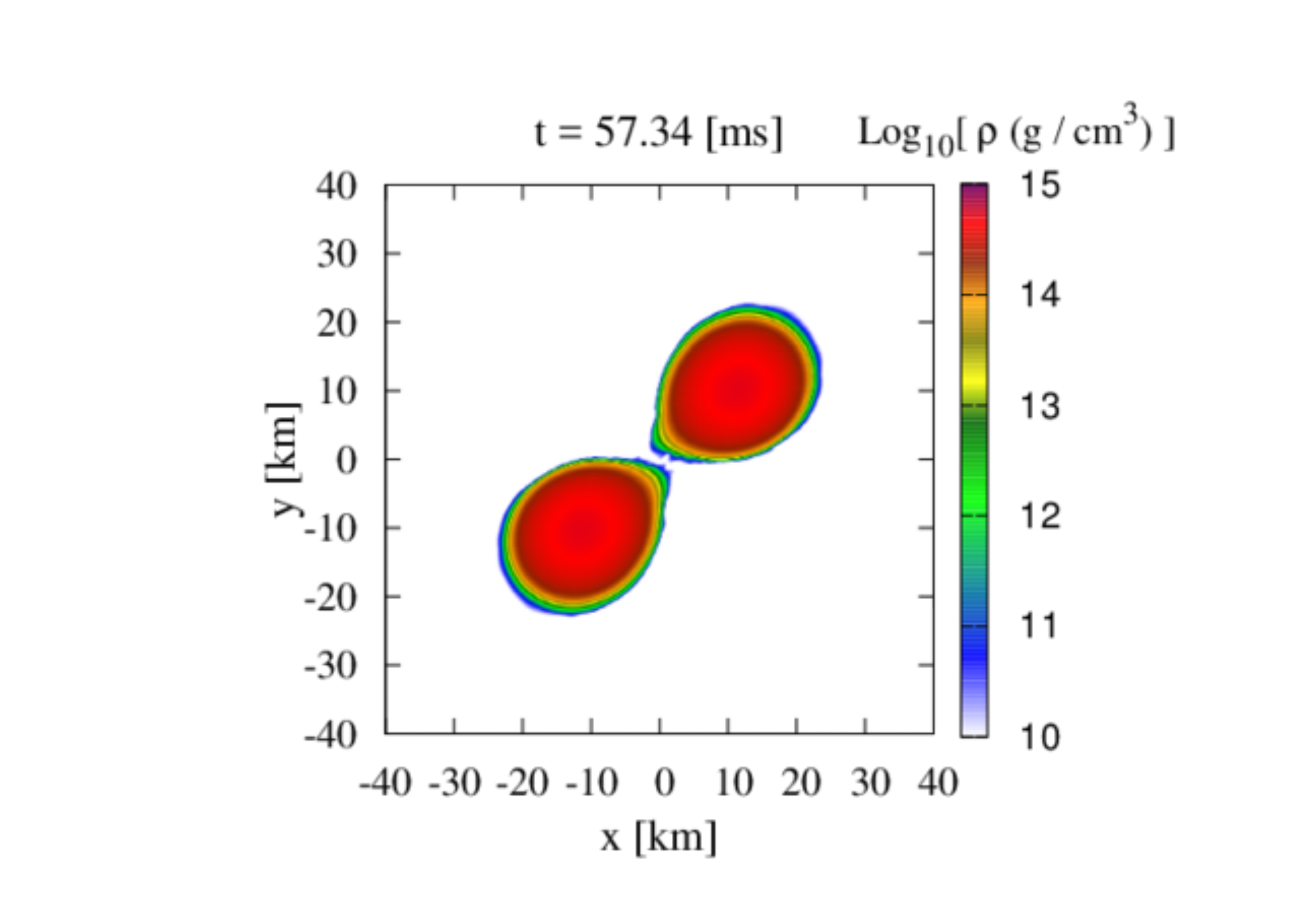}
\hspace{-2cm}\includegraphics[width=50mm,angle=270]{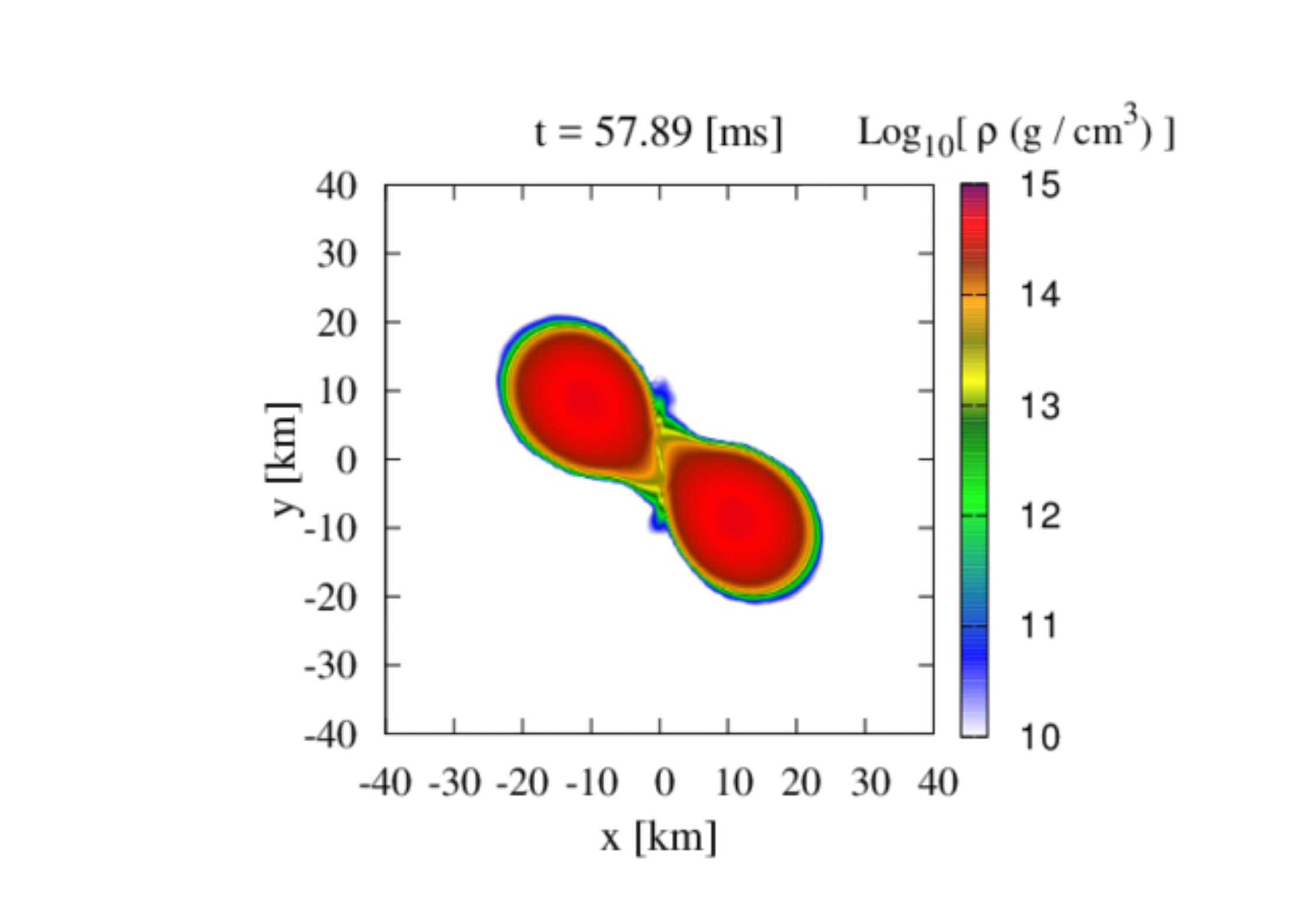}
\hspace{-2cm}\includegraphics[width=50mm,angle=270]{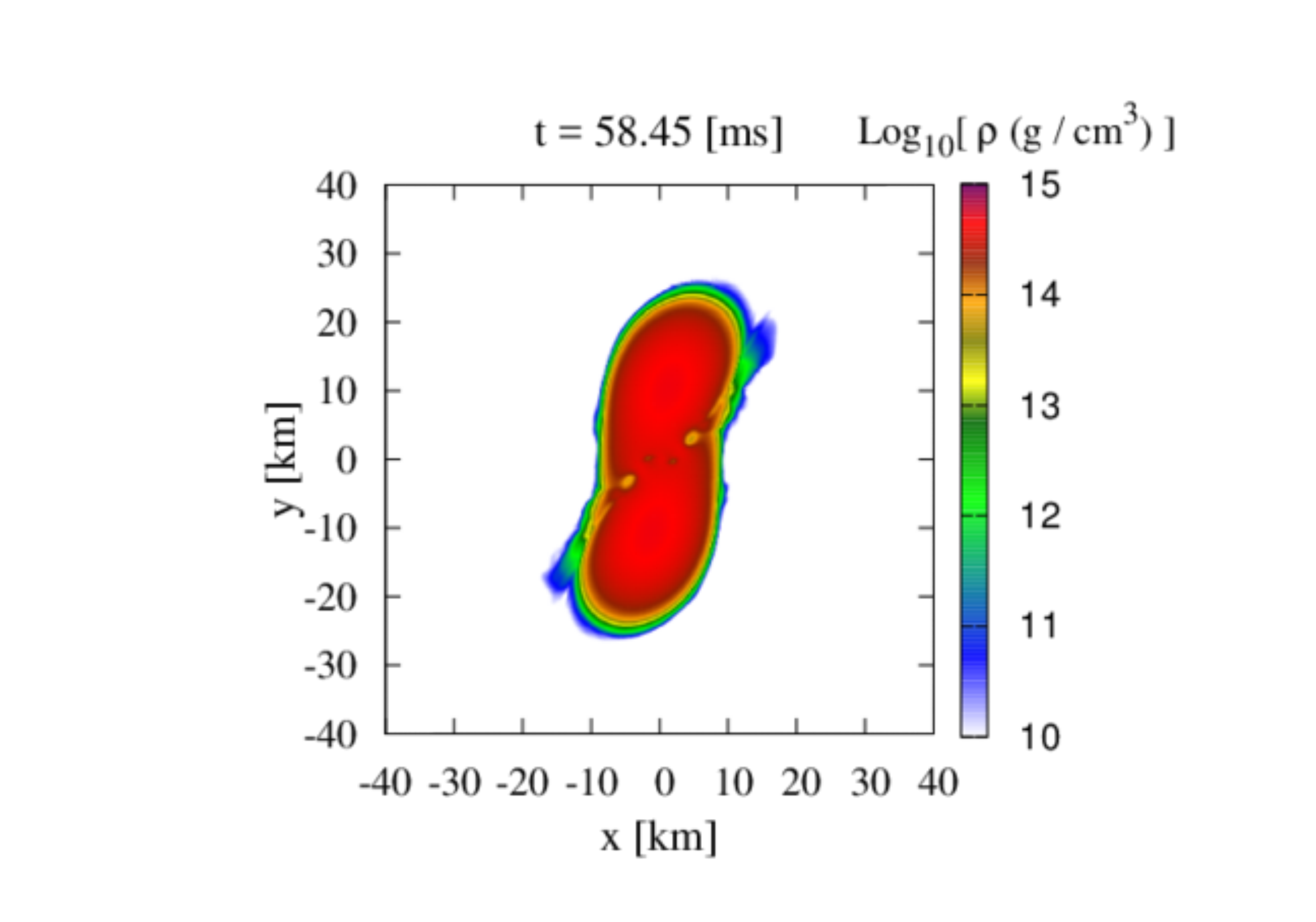}
\caption{Density profiles on the equatorial plane for the model
  15H135-135 with $N=182$ at $t=t_{\rm peak}-1.10\,{\rm ms}$ (left), $t_{\rm
    peak}-0.55\,{\rm ms}$ (middle), and $t_{\rm peak}$ (right). 
\label{fig7}}
\end{center}
\end{figure*}

\begin{table}[t]
\centering
\caption{\label{tab3} $\Phi_{\rm NR}-\Phi_{\rm TEOB}$ at $t=t_{\rm
    peak}$ in units of radian for the equal-mass models with $N=182$
  and 150, and for the unequal-mass models with $N=150$.  }
\begin{tabular}{cccc}
\hline\hline
~~EOS~~ & ~$(\eta, N)=(0.250,182)$~ & ~$(0.250,150)$~ &~$(0.247,150)$~
\\ \hline
B    & 0.1 & 0.3  & 0.2 \\
HB   & 0.3 & 0.6  & 0.4 \\
H    & 0.7 & 0.9  & 0.7 \\
125H & 1.0 & 1.1  & 1.1 \\
15H  & 1.3 & 1.3  & 1.3 \\
\hline\hline
\end{tabular}
\end{table}

In Figs.~\ref{fig5} and \ref{fig6}, we compare the NR waveforms
with the TEOB waveforms. In Table~\ref{tab3}, we also list $\Phi_{\rm
  NR}-\Phi_{\rm TEOB}$ at the moment that the gravitational-wave
amplitude reaches the peak (referred to as $t_{\rm peak}$ in the
following).  Figures~\ref{fig5} and \ref{fig6} show that up to $t_{\rm
  peak}-3$\,ms, the TEOB waveforms well reproduce the NR 
waveforms irrespective of the EOS and mass ratio: In particular for
the models for which the compactness is large and the tidal
deformability is small (e.g., for B EOS), the agreement is quite
good even at $t=t_{\rm peak}-1$\,ms, and the disagreement in the
gravitational-wave phase is within 0.3\,rad up to $t_{\rm peak}$, 
which is within the uncertainty due to the phase error.

On the other hand, for the models for which the compactness is
relatively small (e.g., for 125H and 15H EOS), agreement between the
NR and TEOB waveforms becomes poor for the last few ms prior
to $t_{\rm peak}$, leading to a phase disagreement of $\agt 1$\,rad, 
which is greater than the uncertainty due to the phase error (see Table~\ref{tab3}). 
The interpretation for this is described as
follows: For these small-compactness models, two neutron stars come
into contact at $t_{\rm cont} \sim t_{\rm peak}-1$\,ms (see the left
panel of Fig.~\ref{fig7}).  Then, after the contact, the tidal
deformation is further enhanced and a dumbbell-like object is formed
(see the middle panel of Fig.~\ref{fig7}).  However, the density peaks 
of the dumbbell-like object are still clearly separated, and hence, the
gravitational waveform has a chirp-type signal although the waveform
is different slightly from the chirp signal from the separated body; that is, the evolution
process of the system is determined by hydrodynamics equations (not
simply by two-body equations of motion) and emission process of
gravitational waves by the dumbbell-like object. As the distance
between two density peaks decreases sufficiently, the
gravitational-wave amplitude eventually reaches the peak (see the
right panel of Fig.~\ref{fig7}), and after the two density peaks
merge, the amplitude significantly decreases.  As mentioned above, for
$t_{\rm cont} \leq t_{\rm ret} \leq t_{\rm peak}$, chirp-type
gravitational waves are emitted from a dumbbell-like object.  However,
the evolution of the object and the resulting gravitational waveforms
cannot be well modeled by the current TEOB formalism because this
stage is beyond the range of its application.

\begin{figure*}[th]
\begin{center}
\includegraphics[width=84mm]{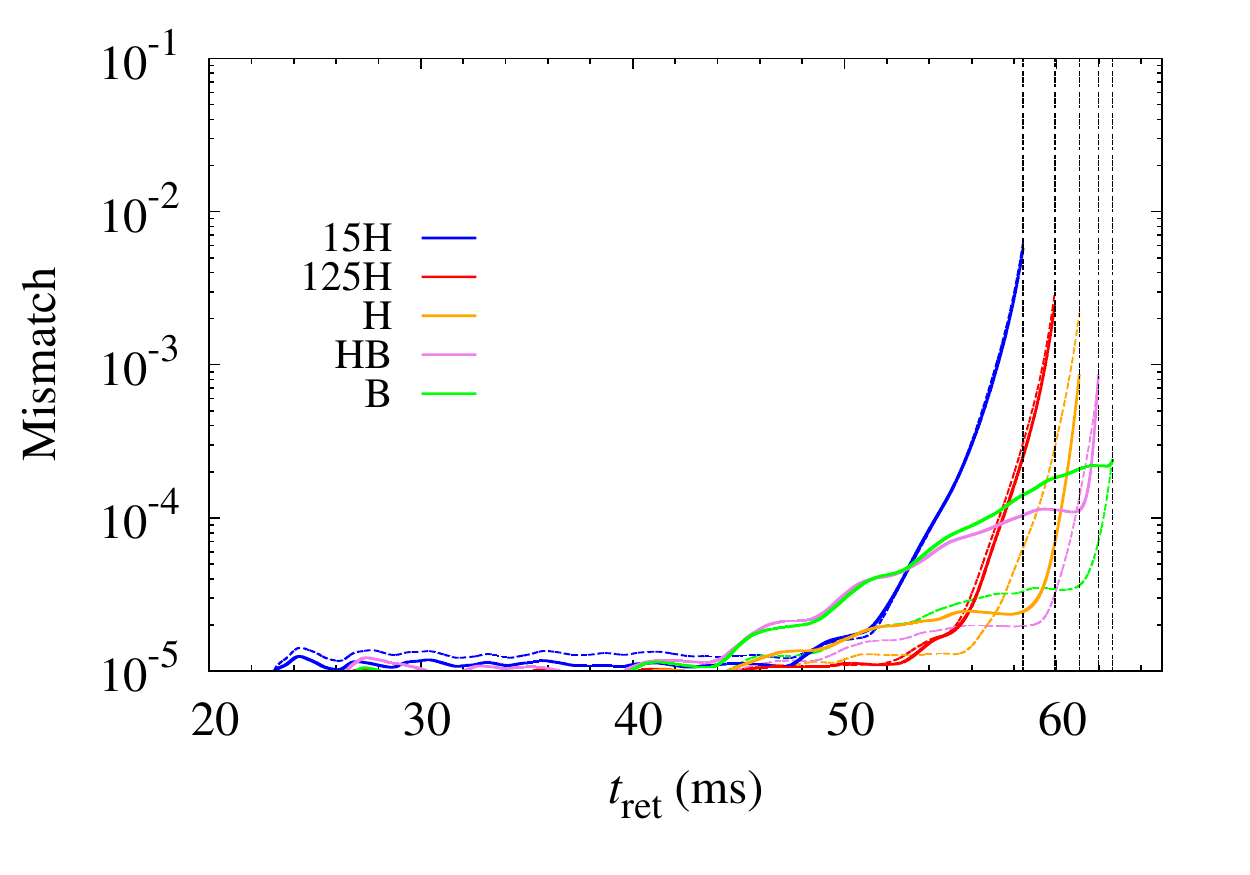}~~~
\includegraphics[width=84mm]{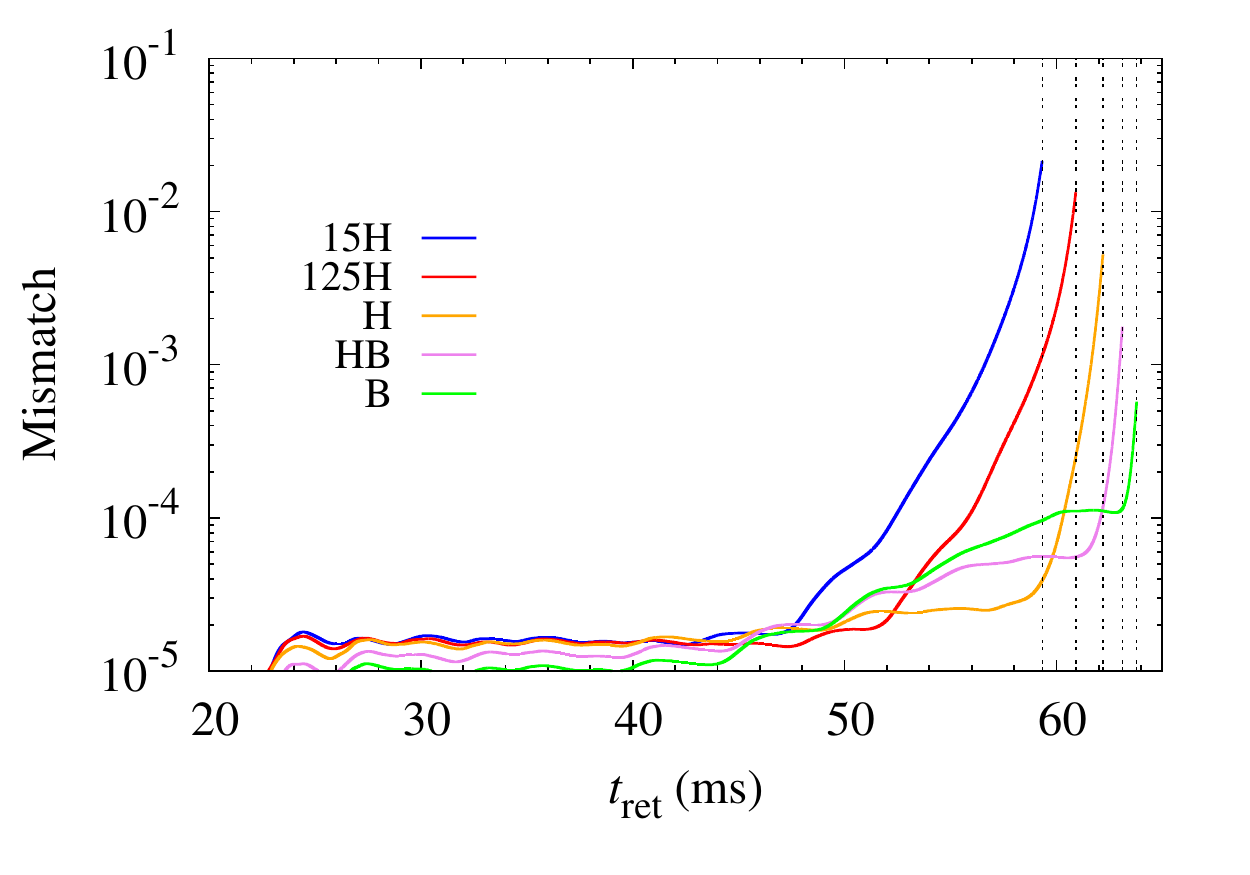}
\vspace{-3mm}
\caption{${\cal M}(t)$ for the equal-mass models (left panel) and for
  the unequal-mass models (right panel).  In the left panel, the thick
  and thin curves denote the results for the numerical waveforms with
  $N=182$ and 150, respectively. In the right panel, the thick curves denote 
  the results for $N=150$. The vertical lines denote $t_{\rm peak}$ for each model.
\label{fig8}}
\end{center}
\end{figure*}

For the larger-compactness models, similar disagreement is found but
only for the short duration because the tidal effects are weak, and
furthermore, $t_{\rm peak} - t_{\rm cont}$ is $\alt 1$\,ms: For such
compact models, the separation between two neutron stars is already
small at $t_{\rm ret}=t_{\rm cont}$, and hence, the time scale of
gravitational radiation reaction is as short as or shorter than the
orbital period.  Therefore, the duration of the stage of the
dumbbell-like configuration is quite short and the disagreement
between the NR and TEOB waveforms are not very remarkable.

Figure~\ref{fig5} and Table~\ref{tab3} show that irrespective of the
numerical data employed, we obtain approximately the same results for
the comparison between the NR and TEOB waveforms. This
reconfirms that for $N\agt 150$, an approximately convergent waveform
(but with the phase error of order 0.1\,rad) can be obtained in our
numerical implementation.

To quantify the disagreement between the NR and TEOB waveforms,
we define a measure of the mismatch by
\beqn
 {\cal M}(t_{\rm ret}):=1-{|(h_{\rm NR}|h_{\rm TEOB})| \over
   (h_{\rm TEOB}|h_{\rm TEOB})^{1/2}(h_{\rm NR}|h_{\rm NR})^{1/2}}, ~~
\eeqn
where $(h_1|h_2)$ is a function of $t_{\rm ret}$ defined 
(without referring to detector noises) 
by
\beq
(h_1|h_2)
:=\int_{t_i}^{t_{\rm ret}} h_1(t_{\rm ret}') h_2^*(t_{\rm ret}') dt_{\rm ret}', 
\eeq
and $h^*$ denotes the complex conjugate of $h$.  Here, $t_i$ is chosen
to be 20\,ms, and $h_{\rm NR}$ and $h_{\rm TEOB}$ denote the
NR and TEOB waveforms, respectively.

Figure~\ref{fig8} plots ${\cal M}(t)$ for all the models with $N \geq
150$ that we consider in this paper. This shows that the degree of the
mismatch steeply increases for the last inspiral orbits, in particular
for the stiff EOS, due to the lack of modeling in the TEOB formalism
as mentioned above. On the other hand, up to $\sim t_{\rm
  peak}-3$\,ms, the match is quite good: ${\cal M}$ is smaller
than $10^{-3}$. This convinces us that the final issue in the TEOB
formalism, in particular for the stiff EOS of large tidal
deformability, is to take into account dynamics and the waveform in the
dumbbell-like object phase.  The phase difference, $\Phi_{\rm
  NR}-\Phi_{\rm TEOB}$, as well as the mismatch, ${\cal M}$, at a
given moment of the retarded time increases nonlinearly with $\tilde
\Lambda$.  This suggests that for the improvement of the TEOB
approach, a new nonlinear term of the tidal deformability is
needed. This point will be discussed in our accompanying
paper~\cite{kawaguchi}.

\section{Summary}

We presented our latest numerical-relativity results of long-term,
high-accuracy simulations for the inspiraling binary neutron stars.
The simulations were performed not only for the equal-mass binaries
but also for the unequal-mass ones.  We showed that if the grid
resolution is high enough (i.e., the neutron-star radii are covered
with the grid of its spacing 60--80\,m), it is possible to obtain a
nearly convergent gravitational waveform (with the phase error of
order 0.1\,rad) from inspiraling binary neutron stars.

By comparing our high-resolution waveforms with the TEOB waveforms, we
find that the TEOB formalism can reproduce accurate waveforms for
binary neutron stars up to $\sim t_{\rm peak}-3$\,ms irrespective of
the neutron star EOS models.  However, it is also found that for
$t_{\rm peak}-3\,{\rm ms} \alt t \alt t_{\rm peak}$ (in particular for
$t_{\rm peak}-1\,{\rm ms} \alt t \alt t_{\rm peak}$), the current TEOB
formalism cannot reproduce the numerical waveforms, in particular for
the binary neutron stars of stiff EOS, and the phase error between the
numerical and TEOB waveforms cannot be negligible as $\agt 1$\,rad at
the amplitude peak for the stiff EOS models. The primary reason for
this is that for such a stage, the evolution of the system cannot be
well reproduced by the current TEOB equation of motion.

Accurate numerical data is crucial for modeling gravitational
waveforms in the frequency domain~\cite{kahn}. Our numerical waveforms
are the most accurate ones among those have been ever derived.  We are
now developing a phenomenological model from our numerical waveforms in the frequency domain, since we
find that the numerical waveforms have a quality enough for the
modeling.  The results will be presented in our accompanying 
paper~\cite{kawaguchi}.

\begin{acknowledgments}

We thank Tanja Hinderer, Ben Lackey, and Andrea Taracchini for helpful
discussions and for telling us the details of the latest TEOB
formalism.  Numerical computation was performed on K computer at AICS
(project numbers hp160211 and hp170230), on Cray XC30 at cfca of
National Astronomical Observatory of Japan, FX10 and Oakforest-PACS at Information
Technology Center of the University of Tokyo, HOKUSAI FX100 at RIKEN,
and on Cray XC40 at Yukawa Institute for Theoretical Physics, Kyoto
University.  This work was supported by Grant-in-Aid for Scientific
Research (24244028, 16H02183, JP16H06342, JP17H01131, 15K05077, 17H06361) 
of JSPS and by a
post-K computer project (Priority issue No.~9) of Japanese MEXT.
Kawaguchi was supported by JSPS overseas research fellowships.

\end{acknowledgments}

\appendix
\section{The effects of residual eccentricity in the numerical waveforms}

\begin{figure*}[t]
\begin{center}
\includegraphics[width=84mm]{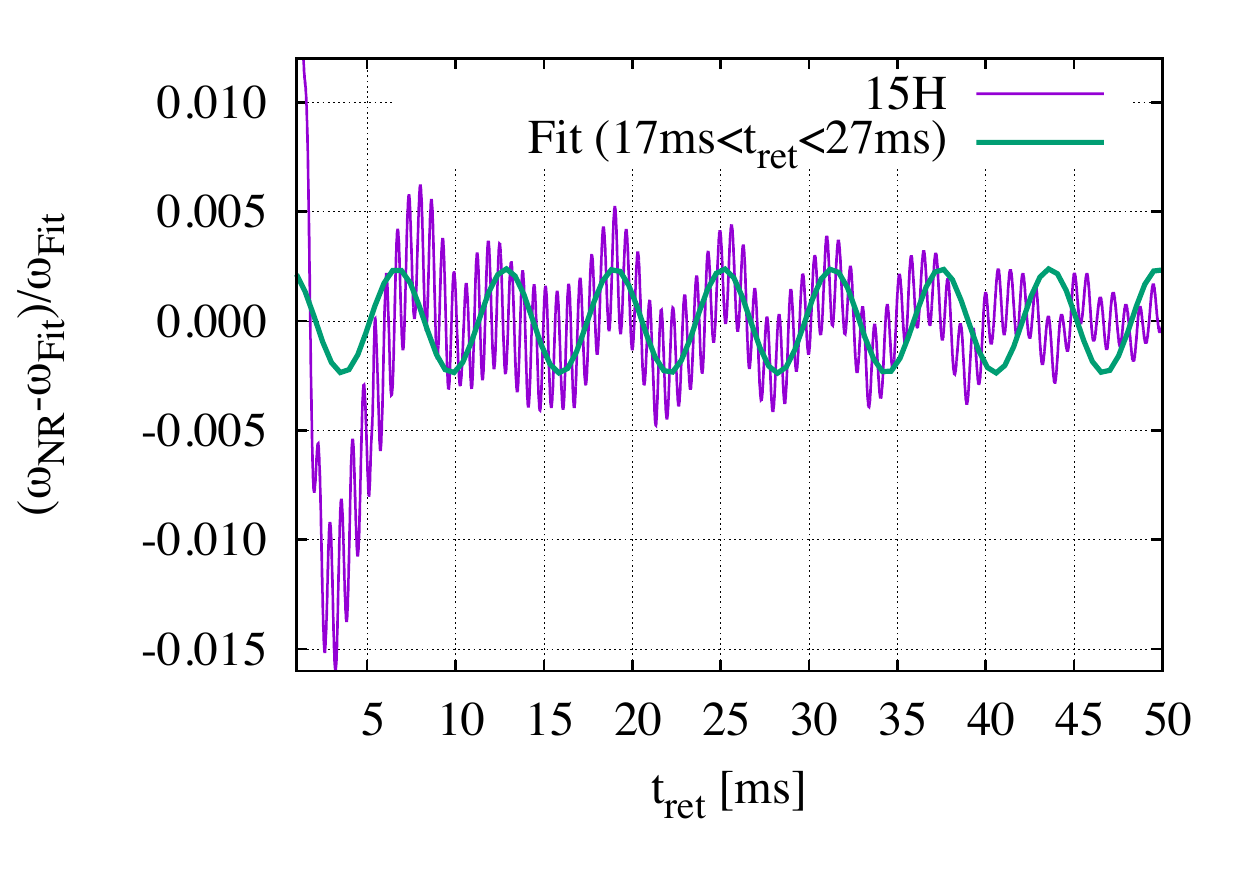}~~
\includegraphics[width=84mm]{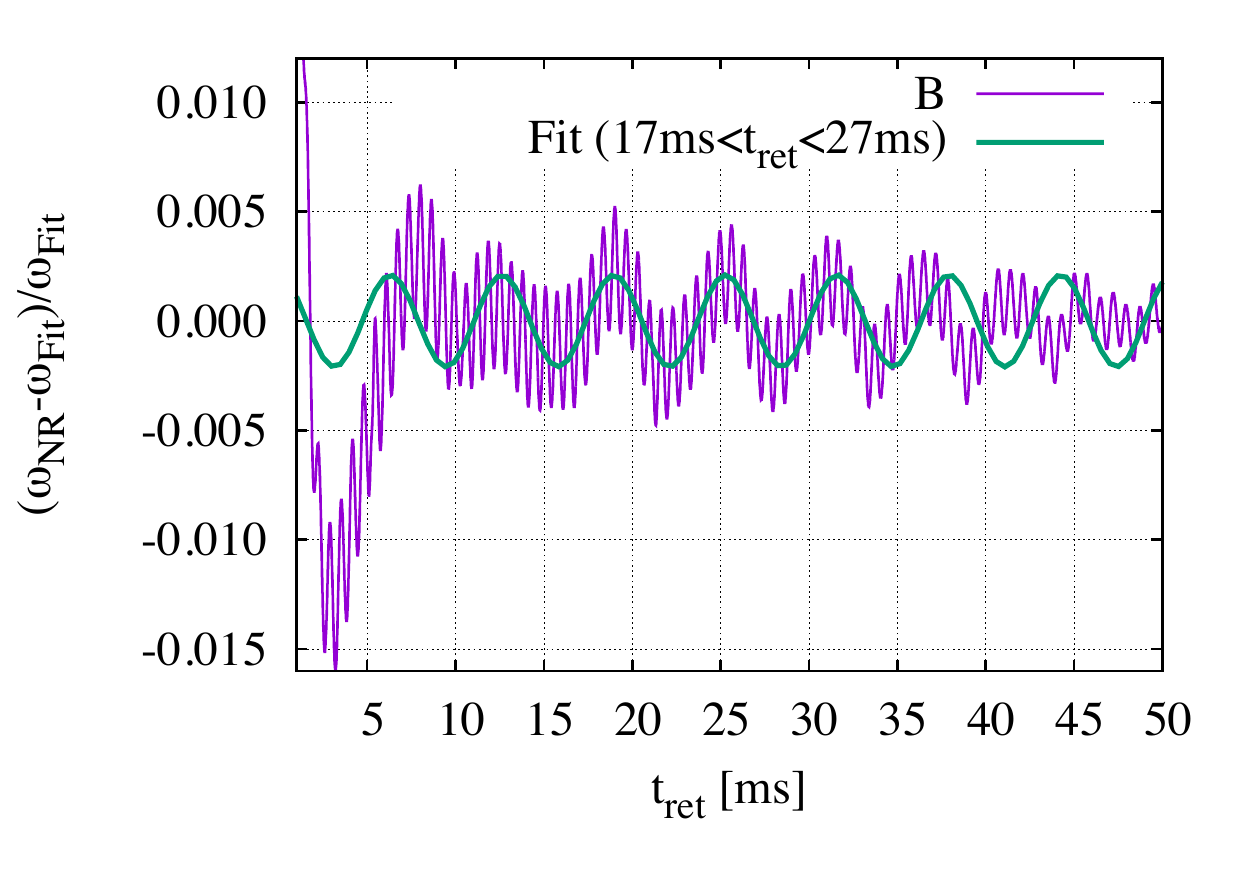}
\vspace{-3mm}
\caption{The modulation in the angular frequency of the numerical
  waveforms for 15H (left panel) and B (right panel) EOS. The
  fitting curves are determined from the data in $17\,{\rm ms} \leq
  t_{\rm ret} \leq 27\,{\rm ms}$ by a sinusoidal function. For these
  fitting curves, the eccentricity of the binary orbits is measured to
  be $e \approx 1.19 \times 10^{-3}$ and $1.05 \times 10^{-3}$ for 15H
  and B EOS, respectively.  }
	\label{fig9}
\end{center}
\end{figure*}

We here demonstrate that the numerical waveforms in the early stage of
numerical simulations are not suitable for the precise analysis
because of the presence of the unphysical and irregular modulation in
the angular frequency in the numerical waveforms, which is likely to
be generated due to an unphysical setting of initial conditions (see
below).

The modulation in the angular frequency of the numerical waveforms is
extracted in the following procedure. First, we fit and subtract the
non-oscillatory (i.e., physical) part of the numerical angular
frequency by employing a function of the form
\begin{align}
\omega_{\rm
  Fit}\left(t\right)=\sum_{n=0,n\neq1}^{7}a_n
\left(t_1-t\right)^{-(n+3)/8}.\label{eq:fit_ome}
\end{align}
The choice of this form is motivated by the Taylor-T3
approximant~\cite{2014LRR....17....2B,2009PhRvD..80h4043B}, and the
coefficients, $t_1$ and $a_n$, are determined by the least-square
fitting procedure.  The fitting is performed for $5\,{\rm ms}\le
t_{\rm ret}\le50\,{\rm ms}$ of the waveforms. We checked that the
result of the fitting depends only weakly on the choice of the time
window.

Figure~\ref{fig9} shows the modulation in the angular frequency
of the numerical waveforms for the equal-mass model with 15H and B EOS with $N=150$. 
The modulation is defined by
\beq
    {\omega_{\rm NR} - \omega_{\rm Fit} \over \omega_{\rm Fit}},
\eeq
where $\omega_{\rm NR}$ is the angular frequency of the
numerical-relativity waveforms.  The fitting curves in this figure are
determined from the data in $17\,{\rm ms} \leq t_{\rm ret} \leq
27\,{\rm ms}$ assuming that the modulation is written as a sinusoidal
function. For $e\ll 1$, the residual eccentricity $e$ of the binary
orbits is related to the amplitude of the modulation in the angular
frequency, $\Delta\omega$, by $\Delta\omega\approx2e\omega$ assuming
that the Newtonian relation is satisfied. For these fitting curves,
the eccentricity of the binary orbits is measured to be $e \approx
1.19 \times 10^{-3}$ and $1.05 \times 10^{-3}$ for 15H and B EOS,
respectively. Figure~\ref{fig9} shows that the eccentricity decreases
with time, and it is $\lesssim 10^{-3}$ for $\gtrsim 20\,{\rm ms}$
irrespective of the EOS and binary mass employed.

Figure~\ref{fig9} shows that in the early part of the evolution with
$t_{\rm ret} \lesssim 15\,{\rm ms}$, the modulation in the angular
frequency behaves in an irregular manner, although that for $t_{\rm
  ret} \agt 15$\,ms exhibits a simple damped-oscillation-like feature
(neglecting its fine structure).  That is, for $t_{\rm ret} \lesssim
15\,{\rm ms}$, the center of the oscillation is not located at
zero. This irregular oscillation causes an irregular error in the
gravitational-wave phase and makes it difficult to perform a careful
comparison between the numerical and TEOB waveforms.  Therefore, we
discard the waveforms in the early stage, and use only the data with 
$t_{\rm ret}\ge 20\,{\rm ms}$ for the comparison in this paper. We
note that even when we discard the data with $t_{\rm ret} \leq
20$\,ms, i.e., first $\sim 6.5-7$ wave cycles, we still have $\agt 22$
cycles in the waveforms up to $t_{\rm peak}$.

Our interpretation for this irregular modulation for the first $\sim
15$\,ms is that burst-like junk radiation is emitted just after the
simulation is started, and it takes about 15\,ms until the system
relaxes to a quasi-stationary state. The junk radiation is caused by
unphysical setting of the initial condition associated with the
so-called conformal flatness approximation for the initial-data
problem (e.g., see chapter 5 of Ref.~\cite{NR2016}).

\section{Phase error due to the finite-radius extraction and violation of the baryon mass conservation}

\begin{figure*}[t]
\begin{center}
\includegraphics[width=84mm]{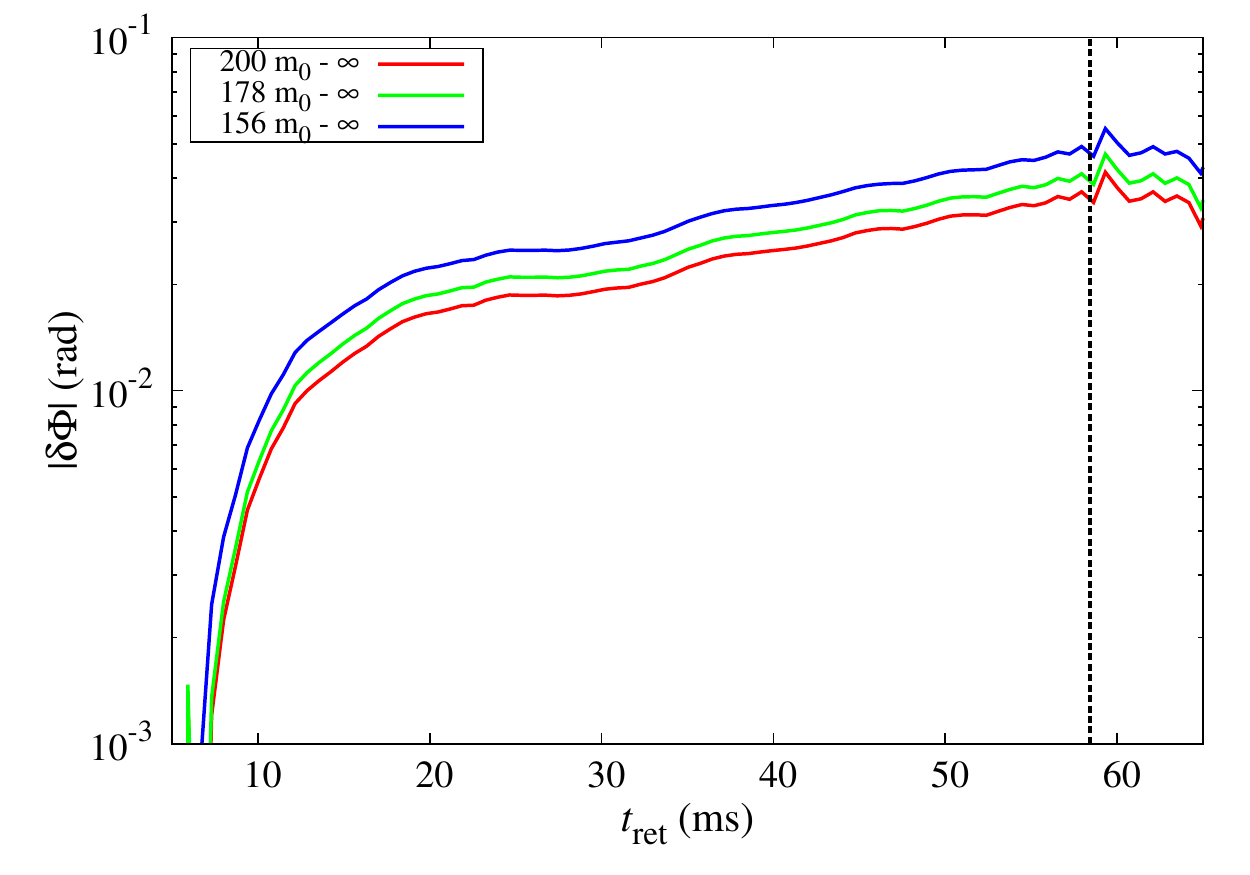}~~
\includegraphics[width=84mm]{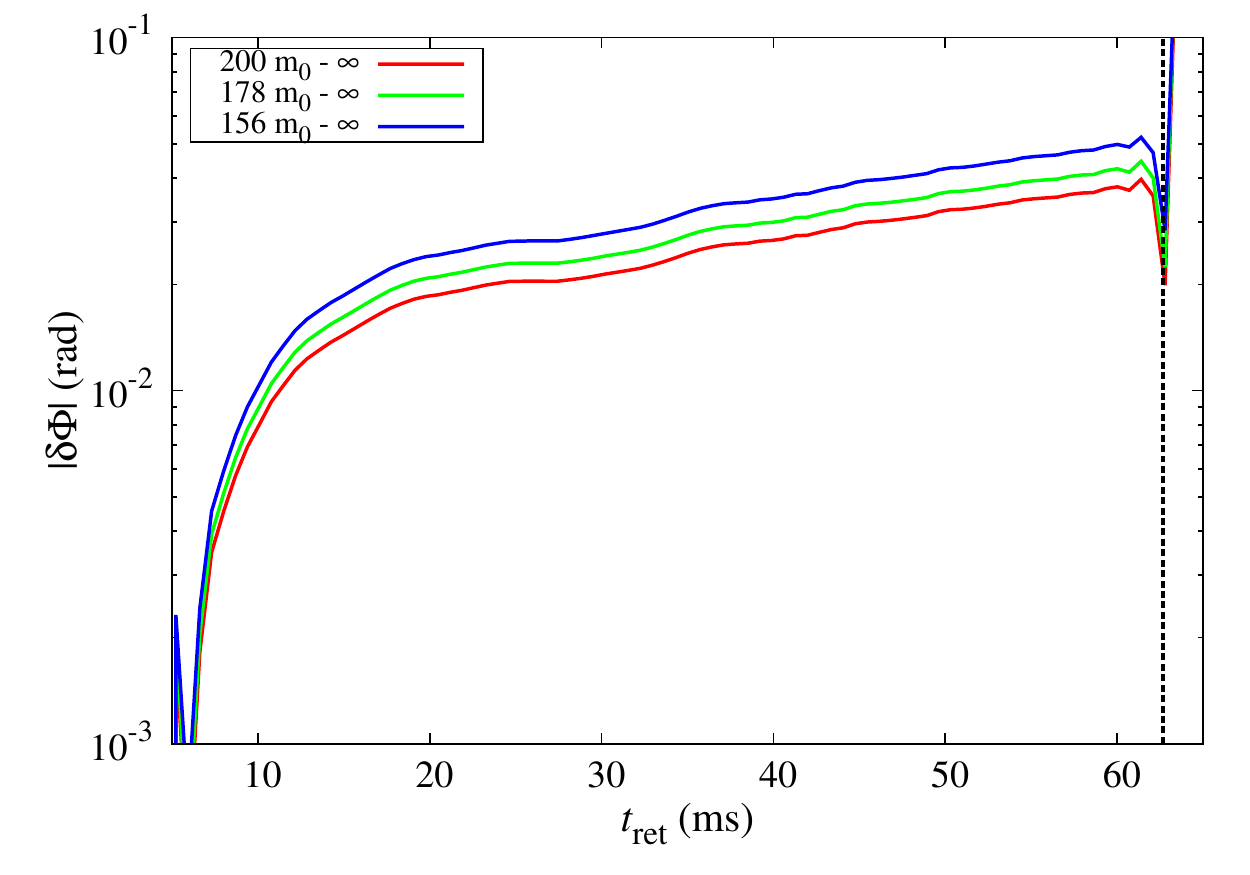}
\vspace{-3mm}
\caption{Phase error due to the finite-radius extraction as a function of the retarded time for 15H EOS (left) and B EOS (right) for 
the equal-mass models with $N=182$. The vertical dotted lines are the peak amplitude time $t_{\rm peak}$ (see the text in detail).\label{fig10}}
\end{center}
\end{figure*}

According to the Nakano's method Eq.~(\ref{eq:nakano}), we extrapolate the gravitational waveforms extracted 
at $r_0/m_0=156$, $178$, and $200$ to infinity. We find that the extrapolated waveforms slightly deviate each other 
due to the finite-radius extraction. By assuming this error falls off as $1/r_0$, we obtain a waveform at infinity from the waveforms 
generated by the Nakano's method. 
Figure~\ref{fig10} plots the phase difference between the waveform generated by the Nakano's method and the waveform at infinity 
as a function of the retarded time. The phase error due to the finite-radius extraction is as small as $\sim 0.01$\,rad up to the 
peak amplitude time and it decreases as a function of the extraction radius. We find a similar trend in the other equal-mass models and the unequal-mass 
models. 

\begin{figure*}[t]
\begin{center}
\includegraphics[width=84mm]{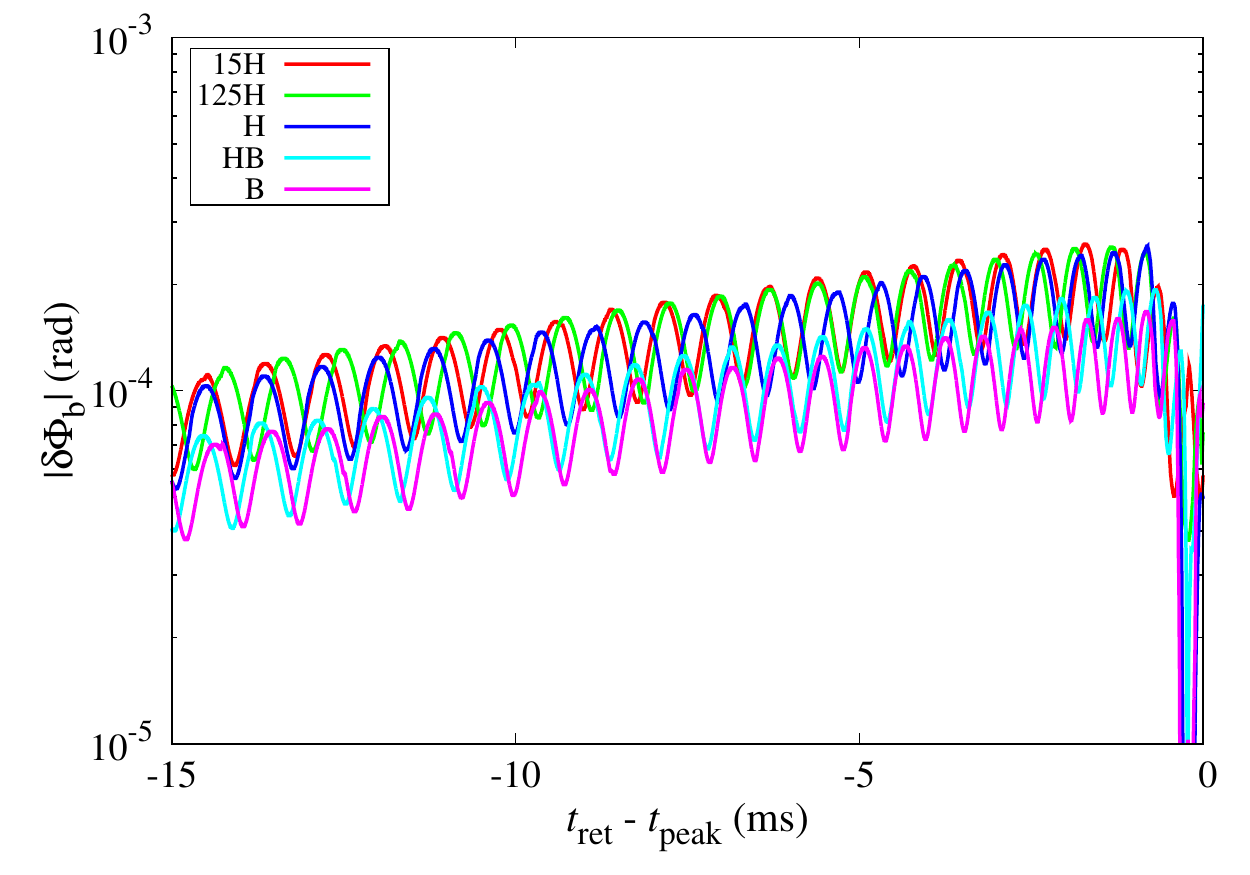}
\vspace{-3mm}
\caption{
Phase error due to the violation of the baryon mass conservation as a function of the retarded time 
for the equal-mass models with $N=182$. The vertical dotted lines show the peak amplitude time of each model. 
\label{fig11}}
\end{center}
\end{figure*}

The baryon mass conservation is slightly violated at the merger because the conservative mesh refinement 
is not implemented in our code~\cite{East:2011aa,Dietrich:2015iva}. Following Ref.~\cite{tim17}, 
we estimate the phase error due to the violation of the baryon mass conservation by
\begin{align}
\delta \Phi_{\rm b} = \omega t_{\rm ret} \frac{\Delta M_{\rm b}}{M_{\rm b}},
\end{align}
where $\omega$ is the angular frequency Eq.~(\ref{eq:omega}) and $M_{\rm b}$ is the baryon mass. 
$\Delta M_{\rm b}$ is the violation of the baryon mass conservation. 
Figure~\ref{fig11} plots the estimated phase error due to the violation of the baryon mass conservation for the equal-mass models with $N=182$. 
This plot shows that the resolution adopted in this work 
is high enough to reduce the error in the baryon mass conservation to $\approx 10^{-4}$ \% up to the peak amplitude time. 
Consequently, the phase error is $\approx O(10^{-4})$ rad at the peak amplitude time irrespective of the models. 
The unequal-mass models show a similar result. 

\begin{thebibliography}{99}
\bibitem{aboutL}
 In Ref.~\cite{Hotoke15}, the size of the finest AMR domains is about $1.15$ times 
as large as the neutron star radii, i.e., $L/2^l=1.15R_{\rm NS}$ with $l=8$. In this work, we conservatively set the finest AMR
 domain size as 
$1.35R_{\rm NS}$.
\end{thebibliography}
\end{document}